\documentstyle[12pt]{article}
\def\ve{{\bf e}}
\def\vm{{\bf m}}
\def\hk{\hat{k}}
\def\hm{\hat{m}}
\def\hn{\hat{n}}
\def\he{\hat{e}}
\def\hl{\hat{l}}

\def\vhm{{\bf \hm}}
\def\vhe{{\bf \he}}
\def\vhk{{\bf \hk}}
\def\vhl{{\bf \hl}}
\def\vhn{{\bf \hn}}

\def\L{\Lambda}
\def\hL{\hat{\Lambda}}
\def\hG{\hat{G}}
\def\hg{\hat{{\cal G}}}
\def\hw{\hat{w}}
\def\vhw{{\bf \hw}}
\def\O{\Omega}
\def\G{{\cal G }}
\def\S{{\cal S }}
\def\T{{\cal T }}
\def\Z{{\cal Z }}

\def\a{\alpha}
\def\b{\beta}

\def\g{\gamma}

\def\bbar{\beta}

\def\P{\Phi}
\def\r{\rho}
\def\l{\lambda}
\def\s{\sigma}

\def\e{\epsilon}

\def\t{\tilde}
\def\implies{\Rightarrow}

\begin{document}

\newcommand{\Dirac}{/\!\!\!\!D}
\newcommand{\inv}[1]{{#1}^{-1}} 

\renewcommand{\theequation}{\thesection.\arabic{equation}}
\newcommand{\beq}{\begin{equation}}
\newcommand{\eeq}[1]{\label{#1}\end{equation}}
\newcommand{\ber}{\begin{eqnarray}}
\newcommand{\eer}[1]{\label{#1}\end{eqnarray}}
\begin{titlepage}
\begin{center}
\hfill hep-th/9502057 \vskip .01in \hfill NYU-TH-94/12/1, IFUM/488/FT
\vskip .01in \hfill CPTH-A347-01/95, RI-12-94
\vskip .4in
{\large\bf S-Duality in N=4 Yang-Mills Theories with General Gauge Groups}
\end{center}
\vskip .4in
\begin{center}
{\large Luciano Girardello}$^\clubsuit$,
{\large Amit Giveon}$^\diamondsuit$,
{\large Massimo Porrati}$^\heartsuit$, and
{\large Alberto Zaffaroni}$^\spadesuit$
\vskip .1in
$\clubsuit$ Dipartimento di Fisica, Universit\`a di Milano, via Celoria 16,
20133 Milano, Italy\footnotemark
\footnotetext{e-mail girardello@vaxmi.mi.infn.it}
\vskip .05in
$\diamondsuit$ Racah Institute of Physics, The Hebrew University, Jerusalem
91904, Israel\footnotemark
\footnotetext{e-mail giveon@vms.huji.ac.il}
\vskip .05in
$\heartsuit$ Department of Physics, NYU, 4 Washington Pl.,
New York, NY 10003, USA\footnotemark
\footnotetext{e-mail porrati@mafalda.physics.nyu.edu}
\vskip .05in
$\spadesuit$ Centre de Physique Theorique, Ecole Polytechnique,
F-91128 Palaiseau CEDEX,
France\footnotemark
\footnotetext{e-mail zaffaron@orphee.polytechnique.fr; Laboratoire Propre
du CNRS UPR A.0014}
\end{center}
\vskip .4in
\begin{center} {\bf ABSTRACT} \end{center}
\begin{quotation}
\noindent
't Hooft construction of free energy, electric and magnetic
fluxes, and of the partition function with twisted boundary
conditions, is extended to the case of $N=4$ supersymmetric Yang-Mills
theories based on arbitrary compact, simple Lie groups.

The transformation of the fluxes and the free energy under S-duality
is presented.
We consider the partition
function of $N=4$ for a particular choice of boundary conditions, and
compute exactly its leading infrared divergence. We verify that this
partition function obeys the transformation laws required by S-duality.
This provides independent evidence in favor of S-duality in $N=4$ theories.
\end{quotation}
\vfill
\end{titlepage}
\eject
\def\baselinestretch{1.2}
\baselineskip 16 pt
\noindent
\section{Introduction}
\setcounter{equation}{0}
Classical Maxwell's equations can be made symmetric in the exchange of the
electric and magnetic fields by introducing a magnetic current, together with
the electric one, i.e., by introducing magnetic monopoles besides electrically
charged particles. In any quantum theory the magnetic charge of the
monopoles, $g_m$, has to obey the Dirac quantization condition, $g_m=4\pi
n/g_e$, where $n$ is an integer and $g_e$ is the electric charge.
In Abelian gauge theories the monopoles correspond to singular gauge
configuration.

On the other hand, there exist non-Abelian theories where
monopoles exist as classical, non-singular configurations of
finite energy; an example is the $SO(3)$ gauge theory, with a scalar
field in the adjoint representation, breaking the symmetry to $U(1)$.
The existence of classical finite-energy monopoles means that there exist
magnetically charged particles in the quantum theory. Their mass spectrum is
in general difficult to compute since its classical value is modified by
perturbative corrections, still, it was conjectured that a theory with
monopoles could be symmetric under the exchange of electrically charged
particles with monopoles, together with $g_e \leftrightarrow
g_m$~\cite{MO}.

A remarkable simplification happens in the $N=4$
supersymmetric Yang-Mills theory. There, the supersymmetry algebra~\cite{WO}
gives an
exact result for the mass of electrically and/or magnetically charged
particles belonging to ``short multiplets'' (supersymmetry multiplets
containing only states of spin not greater than one). It turns out that if a
short multiplet has $p$ units of electric charge, and $q$ units of magnetic
charge, its square mass is equal to a universal constant times
$p^2g_e^2 + q^2g_m^2$. The existence of magnetic monopoles, with $p=0$,
$q=1$, besides the electrically charged ones, and the fact that their
mass is symmetric under the exchange $g_m \leftrightarrow g_e$, $p
\leftrightarrow q$ gave rise to the hypothesis that $N=4$ super Yang-Mills
is the most likely theory to verify the Montonen-Olive conjecture~\cite{WO,O}.

This conjecture can be generalized to arbitrary gauge groups following
ref.~\cite{GNO}:
in general, the duality $g\rightarrow 4\pi/g$
transforms the gauge group $G$ into another group (the magnetic group)
$\hat{G}$, whose weight lattice is dual to the weight lattice of $G$.

Arguments in favor of electric-magnetic duality for the S-matrix of
$N=4$ where given in~\cite{R}, thereby making it plausible that
$g\rightarrow 4\pi/g$ is a true property of the theory and not only of its
mass spectrum. It was also realized in lattice calculations~\cite{CR}
that in the presence of a theta term,
electric-magnetic duality and theta shifts, $\theta \rightarrow \theta + 2\pi
$, could be combined into an $SL(2,Z)$ group acting on the complex parameter
\beq
S= {4\pi\over g^2} + i {\theta\over 2\pi}
\eeq{i1}
by fractional transformations
\beq
iS \rightarrow {a iS + b \over c iS + d}, \;\;\; a,b,c,d\;\mbox{integers},\;
ad-bc=1.
\eeq{i2}
This enlarged symmetry, called S-duality, maps states with, say, $p=1$, $q=0$
(elementary states, electrically charged), into states with any $p$ and $q$
relatively prime. Recently, Sen~\cite{Sen}
has constructed the previously unknown
states with $p=2$, $q=1$; this result gives important new evidence for the
S-duality conjecture.

S-duality includes the original strong-weak coupling duality
$g\leftrightarrow 4\pi/g$ at zero theta angle.
In~\cite{SW} it was shown that this duality can be
used to compute exactly the low-energy effective action of $N=2$
asymptotically free theories, even
though $N=2$ is not explicitly S-dual. These arguments have been extended to
phenomenologically interesting $N=1$ models~\cite{IS} and provide a new
way of understanding the physics of gauge theories at strong coupling.

It is possible that S-duality is also
a symmetry of strings theory~\cite{FIL}. Evidence for this conjecture
has been given, by different methods, in~\cite{5bra,SS,Gau}. If true,
this would be extremely important
since S-duality is fundamentally non-perturbative (it relates strong and weak
coupling) and does not hold order by order in the string loop expansion;
thus, it may give a way of studying the non-perturbative dynamics of
strings. Together with T-duality (see~\cite{GPR} for a review)
it may also be a
part of a much larger (infinite dimensional) stringy symmetry~\cite{Schwarz}.

Moreover, if S-duality is a fundamental symmetry of string theory, it will
explain its appearance in gauge theories. That is because the
latter can be
derived as the $\a'\to 0$ limit of string theory ($\a'$ is the inverse
string tension), and S-duality should hold
order by order in the $\a'$ expansion.

Most of the evidence for S-duality is classical or semiclassical in nature,
and its validity for the full quantum theory relies on the existence of
non-renormalization theorems. For this reason, strong-coupling tests
of the conjecture are difficult. A strong-coupling test of
S-duality was given in~\cite{VW}, where a topological, twisted version of
$N=4$ was defined, and the corresponding partition function  computed on
various manifolds was found to be S-dual.

In~\cite{GGPZ} we proposed an ansatz for the thermodynamical free energy of
$N=4$, $SU(2)$ Yang-Mills theory in a finite-size box,
and analyzed its transformation properties under
S-duality. This free energy depends on the non-Abelian generalization of
the electric and magnetic fluxes, defined long ago by 't Hooft~\cite{tH}.
In~\cite{GGPZ} we computed the free energy in the weak coupling limit,
and then proposed a possible extension to any $S$ by requiring S-duality.

In this paper, we define an appropriate
partition function (and the corresponding free energy)
for $N=4$ based on an {\em arbitrary} compact, simple Lie group $G$ (the
generalization to any compact group is obvious). Moreover,
we {\em compute} the leading infrared divergence term of
the partition function, by a path-integral formulation.
This free-energy turns out to be identical with the one proposed
in~\cite{GGPZ}, for the case $G=SU(2)$, and turns out to be S-dual for
any gauge group. Namely,
the partition function transforms as demanded by S-duality.

The electric and magnetic fluxes
transform appropriately into other electric and magnetic fluxes for any
element of the S-duality group, $SL(2,Z)$. However, when $G$ is
non-simply laced, some elements of $SL(2,Z)$ transform ``legal'' fluxes
into ``illegal'' fluxes. For those believing that $SL(2,Z)$
S-duality originates as a
fundamental symmetry of string theory, these might be ``good news.'' This is
because non-simply laced gauge groups are impossible in the $D=4$, $N=4$
heterotic string compactifications (whose $\a'\to 0$ limit gives rise to
$N=4$ Yang-Mills theories in $D=4$).

The paper is organized as follows. In Section 2, we review the construction
of $N=4$ super Yang-Mills with $\theta$ term.
Section 3 extends 't Hooft's construction
of the free energy of a flux to any compact gauge group
(the original construction
was given only for $SU(N)$). The new feature that appears for non-simply
laced groups is that the gauge algebrae of $G$ and $\hat{G}$ are different,
and get interchanged under electric-magnetic duality.
Section 4 contains a path-integral computation
of the leading infrared divergence of the partition function and of the
corresponding free energies.
This can be carried out exactly with an appropriate
choice of the boundary conditions and gauge fixing,
and results in an explicit formula which
can be used to test S-duality.
In Section 5, we show that the free energy defined in the previous Section
obeys factorization,
't Hooft duality, and
properly accounts for the shift in electric charge that occurs when the
theta angle is changed  (Witten's phenomenon~\cite{W}). Section 6
contains the explicit transformation laws that the free energies of {\em
any} S-dual theory must obey.
These transformation laws arise independently
of any particular realization of the free energy, and only depend on the
assumption of S-duality symmetry. Still in
Section 6, we verify that the free energies obtained in Section 4 do transform
as prescribed by S-duality. Section 7 contains a summary and some final
remarks. In Appendix A, we set up the
notation needed to carry on the construction of the free energy for
arbitrary simple Lie groups. Appendix B contains an alternative computation
of the partition functions, performed in the Hamiltonian formalism.
The computation of its leading infrared divergence can be carried out
completely at weak coupling, and it
gives rise to the same leading infrared divergence of the path integral
computation, thus providing another check of our formulae.
\section{$N=4$ Supersymmetric Yang-Mills Theories}
\setcounter{equation}{0}
An $N=4$ super Yang-Mills theory is completely determined by the gauge
group, $G$. The fields in the Lagrangian ($L$) form a supermultiplet:
\ber
\P=(A_{\mu}^a,\,\,\, \l_I^a,\,\,\, \phi_{IJ}^a)& & \nonumber \\
\mu=1,2,3,4, \qquad I,J=1,2,3,4, & & \quad a=1,...,d, \quad d={\rm dim}\, G.
\eer{AlP}
All the fields in (\ref{AlP}) are in the adjoint representation of the
gauge group. The supermultiplet contains a gauge field (spin-1),
$A_{\mu}^a$ ($\mu$ is a space-time vector index and $a$ is a group index of
the adjoint representation),
four Weyl spinors (spin-1/2), $\l_I^a$ ($I$ is the so-called "extension
index,'' in the 4 of $SU(4)$, representing the four supersymmetry charges),
and six scalars (spin-0), $\phi_{IJ}^a$, which obey the condition:
$2\phi_{IJ}^a=\e_{IJKL}(\phi_{KL}^a)^*$.
The Lagrangian, which can be obtained most easily by dimensional reduction
of the $N=1$ Yang-Mills theory in 10 dimensions~\cite{BSS}, takes the form
\ber
L &=& \frac{1}{4\pi}{\rm Re}\, S \Big[{1\over 2}F_{\mu\nu}^aF^{a\,\mu\nu}
+ \bar{\lambda}^{a\, I}\Dirac \lambda_I^a
+ D_{\mu}\phi^{a\, IJ}D^{\mu}\phi^a_{IJ}
\nonumber \\ & &
+ f_{abc}\bar{\lambda}^{a\, I}\phi_{IJ}^b\lambda^{c\, J} +
f_{abc}f_{ade}\phi^b_{IJ}\phi^{c\,JK} \phi^d_{KL}\phi^{e\,LI}\Big]
\nonumber\\ & &
-i \frac{1}{8\pi}{\rm Im}\, S F_{\mu\nu}^a\tilde{F}^{a\,\mu\nu}.
\eer{L}
Here $\phi^{a\, IJ}\equiv (\phi^a_{IJ})^*$, $2\tilde{F}^{a\,\mu\nu}=
\e^{\mu\nu\s\r}F^a_{\s\r}$,
and $f_{abc}$ are the structure
constants of $G$. The Cartan-Killing metric is $\delta_{ab}$.
By eq.~(\ref{L}), one finds the relation between $S$, the coupling
constant $g$, and theta angle $\theta$:
\beq
S=\a^{-1}+ia,\qquad \a={g^2\over 4\pi},\qquad a={\theta\over 2\pi}.
\eeq{Saa}
The first three terms in $L$ are the kinetic terms of the gauge fields,
spinors and scalars, respectively, with covariant derivatives due to gauge
symmetry. The fourth term in $L$ is the Yukawa interaction, and the fifth
term is the scalar potential. The latter has flat directions when the scalar
fields get Vacum Expectation Values (VEVs) in the Cartan Sub-Algebra (CSA).
In $N=4$ super Yang-Mills theory, the potential does not receive
perturbative quantum corrections \cite{ro,seiberg}.
Indeed, $N=4$ is so rigid as to forbid
even non-perturbative corrections~\cite{seiberg}. This result agrees with
explicit instanton calculations, as discussed in the second paper of
ref.~\cite{seiberg}.
The last term in $L$ is a topological invariant.

Our aim is to find appropriate gauge invariant quantities which are simple
enough to be calculable, yet non-trivial, i.e. carrying
some dynamical information about the theory. One possibility is to follow
the strategy outlined by 't Hooft in~\cite{tH}, where the non-Abelian
equivalent of the electric and magnetic fluxes are defined. Ref.~\cite{tH}
also contains the definition of the free energies of electric and magnetic
fluxes, in terms of appropriate functional integrals with twisted boundary
conditions, when the gauge theory contains only fields in the adjoint of the
gauge group, and the group itself is $SU(N)$. The restriction to the adjoint
representation is automatic in $N=4$, since all matter fields belong to the
same multiplet and, therefore,
their representation is the same as that of the gauge
vectors. On the other hand, we would like to generalize 't
Hooft's prescription to arbitrary gauge groups.

Before doing this in the next Section, let us
recall that 't Hooft free energies may be used to classify the phases of
pure Yang-Mills. In our case the presence of scalars, and of flat
directions in the scalar potential, alters significantly the picture.
Indeed, by choosing for instance periodic boundary conditions for the
scalars one ends up evaluating the free energy in a mixed phase. Thus 't
Hooft boxes may not be enough to distinguish different dynamical
realizations of the gauge symmetry.

\section{'t Hooft Box with Twisted Boundary Conditions, the Free Energy,
and the S-Duality Conjecture}
\setcounter{equation}{0}
The notations and normalizations we shall use in this work are
introduced in Appendix A. The reader interested in the
algebraic details might find it useful to read Appendix A at this point.

We now want to evaluate the Euclidean functional integral in a box of sides
$a_{\mu}=(a_1,a_2,a_3,a_4)$ and with twisted boundary conditions
\beq
W[\vhk,\vhm]=\int [dA^a_\mu\, d\lambda^a_I\, d\phi_{IJ}^a] \exp (-\int
d^4x L),
\eeq{W}
where the Lagrangian $L$ is given in (\ref{L}), and
\ber
\vhk\equiv (\hk_1,\hk_2,\hk_3),\qquad \vhm\equiv
(\hm_1,\hm_2,\hm_3),\nonumber\\
\hk_i, \hm_j \in {\hL_W \over \hL_R} \simeq C.
\eer{kmC}
Here $i,j=1,2,3$ label the spatial coordinates, and $\hk_i, \hm_j$ are
vectors in the dual weight lattice modulo the root lattice, for any $i$ and
$j$, namely, each represents an element of the center $C$ (see
Appendix A, in particular, eqs. (\ref{hLR}),(\ref{hLW}) and (\ref{Center})).
The center elements $\hk_i, \hm_j$ are defined through the boundary
conditions of the fields, which read
\beq
\P(x+a_{\mu}e_{\mu})=(-)^F\O_{\mu}(x)\P(x),
\eeq{PO}
where $e_{\mu}$ is a unit vector in the $\mu$ direction, and repeated
indices are not summed.
$\P$ and $\O\P$ denote generically a field of
the supermultiplet (\ref{AlP})
and its gauge transform under $\O$, respectively; $F$ is the fermion number:
$F=0$ on bosons and $F=1$ on fermions.
In other words, the boundary conditions for all bosonic fields are periodic,
up to a gauge transformation, whereas fermions are {\em antiperiodic}, up to a
gauge transformation.
Going from $x$ to $x+a_{\nu}e_{\nu}+a_{\mu}e_{\mu}$, $\mu\neq\nu$,
in two different ways -- either in the $\nu$ direction first and then in the
$\mu$ direction or vice-versa -- implies the consistency conditions:
\ber
\O_{\mu}(x+a_{\nu}e_{\nu})\O_{\nu}(x) &=&
\O_{\nu}(x+a_{\mu}e_{\mu})\O_{\mu}(x)z_{\mu\nu},
\nonumber\\
z_{\mu\nu}&\in& C.
\eer{OOOO}
The relation (\ref{OOOO}) is consistent, because a constant gauge
transformation in the center acts trivially on fields in the adjoint
representation. The constant center elements $z_{\mu\nu}$ can be written
explicitly as
\beq
z_{\mu\nu}\equiv z_{\hw_{\mu\nu}}=e^{2\pi i\hw_{\mu\nu}\cdot T} , \qquad
\hw_{\mu\nu}\in {\hL_W\over \hL_R}.
\eeq{zmn}
Equation (\ref{OOOO}) implies that
\beq
\O_{\nu}(a_{\mu})\O_{\mu}(0)=\O_{\mu}(a_{\nu})\O_{\nu}(0)z_{\mu\nu}^{-1}=
\O_{\mu}(a_{\nu})\O_{\nu}(0)z_{\nu\mu},
\eeq{stam}
and from (\ref{stam}) and (\ref{zmn}) we learn that
\beq
z_{\nu\mu}=z_{\mu\nu}^{-1}\,\, \Leftrightarrow\,\, \hw_{\nu\mu}=-\hw_{\mu\nu}.
\eeq{ww}
Thus, $\hw_{\mu\nu}$ is antisymmetric in the space-time indices.

The elements $\hm_i$ in $W[\vhk,\vhm]$ (\ref{W}) are defined by the twists
in the spatial directions:
\beq
\hm_i\equiv {1\over 2}\e_{ijk}\hw_{jk},\qquad i,j,k=1,2,3.
\eeq{m}
$\hm_i$ are interpreted as non-Abelian ``magnetic fluxes'' \cite{tH}.
The elements $\hk_i$ in $W[\vhk,\vhm]$ are defined by the twists in the
time and space directions:
\beq
\hk_i\equiv \hw_{4i}, \qquad i=1,2,3.
\eeq{k}
$\hk_i$ are interpreted as the dual ``electric fluxes.'' The non-Abelian
electric fluxes, $e_i\in \L_W/\L_R$, are linked to $\hk_i$ by the equation
\cite{tH}:
\beq
e^{-\b F[\ve,\vhm]}={1\over N^3}\sum_{\vhk\in (\hL_W/\hL_R)^3}
e^{2\pi i \ve\cdot \vhk} W[\vhk,\vhm].
\eeq{F}
Here $\b\equiv a_4$ is the inverse temperature, $F[\ve,\vhm]$ is the
free energy of a configuration with electric flux $\ve$ and magnetic flux
$\vhm$:
\beq
\ve=(e_1,e_2,e_3),\,\,\, e_i\in \L_W/\L_R,\qquad  \vhm=(\hm_1,\hm_2,\hm_3),
\,\,\, \hm_i\in \hL_W/\hL_R,\qquad
\ve\cdot\vhk\equiv \sum_{i=1}^3 e_i\cdot\hk_i,
\eeq{vevhm}
and
\beq
N=Order(C\simeq \hL_W/\hL_R).
\eeq{NofC}
(Recall that the detailed setting of the notations and normalizations is
given in Appendix A.) Equation (\ref{F}) means that
the free energy is given by the discrete Fourier transform of the
functional integrals $W[\vhk,\vhm]$.

The S-duality conjecture is that under the inversion  of the
complexified coupling constant $S$ in (\ref{Saa}), the free energies
transform as:
\beq
F[\ve,\vhm,1/S,\G]=F[\vhm,-\ve,S,\hg],
\eeq{StoSinv}
while under a theta shift of $S$, they transform as:
\beq
F[\ve,\vhm,S+i,\G]=F[\ve+\vhm,\vhm,S,\G].
\eeq{Sshift}
The transformations $S\to 1/S$ and $S\to S+i$ generate the S-duality group,
isomorphic to $SL(2,Z)$. We should remark that these transformation laws
are expected to hold in any S-duality symmetric theory, independently of
the specific form of the partition function. We will discuss S-duality in
$N=4$ Yang-Mills theories in detail in Section 6. But first, in the next
Section, we will isolate the leading infrared-divergent contribution to
$W[\vhk,\vhm]$ in (\ref{W}), and evaluate the corresponding free energy.

\section{Computing the Free Energy}
\setcounter{equation}{0}
\subsection{Computing the Twisted Functional Integral}
The twisted functional integral~(\ref{W}) is in general too difficult to
compute. Moreover, it is plagued by infrared divergences, even in a finite
box. They are due to the integration of the scalar zero-modes over the
flat directions of the $N=4$ potential\footnotemark.
\footnotetext{Our realization of eq.~(\ref{OOOO}), given in eq.~(\ref{m10}),
implies that the boundary conditions of the scalar fields belonging to the
Cartan subalgebra are strictly periodic. This means that one must integrate
over all their zero modes.}
Fortunately, these two difficulties cure each other, in the  sense that
the most divergent term in eq.~(\ref{W}) can be computed exactly, up to a
flux-independent constant.
This proceeds as follows: in a box of 3-$d$ volume $V$ and at temperature
$\beta$, we introduce an appropriate infrared regulator
$M\beta V$, \footnotemark
\footnotetext{The regulator has negative mass dimension: [mass]$^{-3}$, thus
the infrared limit is $M\rightarrow \infty$.}
independent of the coupling constant and theta angle, which cuts off the
integral over the scalar zero modes. Then one finds an expansion
in  $M$ for the partition function
\beq
W[\vhk,\vhm]=M^n\{w[\vhk,\vhm] + O(M^{-\e})\},\;\;\;\; \e>0,\;\; n>0.
\eeq{m1}
If the conjectured S-duality holds for $W[\vhk,\vhm]$, it should also hold
order by order in the $M$ expansion and, in particular,
for the coefficient of its leading infrared divergence, $w[\vhk,\vhm]$. Thus
we find a new test of S-duality. As we shall see below, $w[\vhk,\vhm]$,
unlike the full functional integral, can be computed exactly, and its
transformation properties under S-duality easily determined.

First of all we must re-examine the standard gauge-fixing procedure and
adapt it to the case of twisted boundary conditions.
The functional integral and boundary conditions are given in
eqs.~(\ref{W}), (\ref{PO}), (\ref{OOOO}). The functional integral
in a given twist sector $\vhk,\vhm$ is extended to all fields
$\Phi(x)$ obeying the boundary conditions~(\ref{PO}), {\em
and to all $\Omega(x)$
obeying the consistency condition}~(\ref{OOOO}), with a given $z_{\mu\nu}$.
Let $F[\Phi]=0$ be a gauge-fixing condition.
Then, following the standard Faddeev-Popov procedure, we
insert the  identity
\beq
1=\int [d\Omega] \delta[F[\O\Phi]]\det {\delta F[\Omega\Phi]\over \delta
\Omega},
\eeq{m2}
into the functional integral~(\ref{W}). The functional integration here
extends to all gauge transformations (periodic {\em and} non-periodic).
Using the gauge invariance of the classical action and integration
measure\footnotemark
\footnotetext{The gauge symmetry of $N=4$ is never anomalous, since all
fermions are in the adjoint representation of the gauge group.}
one finds
\ber
W[\vhk,\vhm]&=&\int [d\Phi][d\Omega]\delta[F[\Omega\Phi]] \det {\delta
F[\Omega\Phi]\over \delta \Omega} \exp(-\int d^4x L)\nonumber \\ &=&
\int[d\Omega]\int'[d\Phi] \delta[F[\Phi]] \det {\delta F[\Phi]\over \delta
\Omega} \exp(-\int d^4x L).
\eer{m3}
The functional integral $\int'[d\Phi]$ extends to all $\Phi$ obeying
$ \Phi(x+a_\mu e_\mu)=(-)^F\Omega'_\mu (x)\Phi(x)$, where
\beq
\Omega'_\mu(x)=\Omega(x+a_\mu e_\mu) \Omega_\mu(x) \Omega^{-1}(x),\;\;\;
\mbox{no sum on }\mu.
\eeq{m4}
It is immediate to prove that $\Omega_\mu'(x)$ obeys the same consistency
conditions as $\Omega_\mu(x)$, namely~(\ref{OOOO}) with the same
$z_{\mu\nu}$.
Since in the functional integral~(\ref{W}) one sums over {\em all}
$\Omega_\mu(x)$ satisfying eq.~(\ref{OOOO}), one may rewrite eq.~(\ref{m3})
as
\beq
W[\vhk,\vhm]={\cal V}\int[d\Phi] \delta[F[\Phi]]
\det {\delta F[\Phi]\over \delta
\Omega} \exp(-\int d^4x L).
\eeq{m5}
Here ${\cal V}$ is the volume of the gauge group,
which is thus factored out in the
standard fashion, and $\Phi$ is integrated over the original range, given by
eqs.~(\ref{PO}), (\ref{OOOO}). Everything works as usual, except that we must
integrate over all transition functions $\Omega_\mu(x)$ satisfying the
consistency conditions~(\ref{OOOO})\footnotemark.
\footnotetext{Alternatively, one may pick up a fixed set of
transition functions, say $\stackrel{o}{\O}$,
and restrict the gauge
integration to transformations which commute with $\stackrel{o}{\O}_\mu(x)$.
We discard this second possibility for reasons explained after
eq.~(\ref{m26}).}

Next, we want to find a gauge fixing which simplifies as much as possible
the computation of the functional integral. Since we expect $N=4$ to be in a
Coulomb phase, it make sense to look for a gauge fixing that selects out the
fields in the Cartan subalgebra of the gauge group. By denoting with
$\alpha$ the roots of Lie~$G$, and with $A^A_\mu(x)$, $A^\alpha_\mu(x)$ the
gauge fields in and outside of the Cartan subalgebra, respectively,
a possible choice is
\ber
(a) && [\partial_\mu -i\alpha_AA^A_\mu(x)]A^\alpha_\mu(x)\equiv D^{Ab.}_\mu
A_\mu^\alpha(x) =0,\;\;\; \alpha\neq 0 \nonumber\\
(b) && \partial_\mu A^A_\mu(x)=0,\;\;\;\; A=1,..,r=\mbox{rank } G.
\eer{m6}
The gauge fixing~(a) reduces the gauge symmetry to $U(1)^r$, whereas~(b) fixes
this Abelian symmetry.
The transition functions obeying the gauge-fixing
condition (a) in~(\ref{m6}) are
\beq
\Omega_\mu(x) =e^{2\pi i \omega_\mu^A(x)T_A}.
\eeq{m7}
These transition functions are elements of the Cartan subgroup.
Indeed, for a generic $A_{\mu}$ in the gauge~(\ref{m6}),
$D_{\mu}^{Ab.}A_{\mu}=0$
together with $D_\mu^{Ab.}\O A_{\mu}=0$ implies $D_\mu^{Ab.}\O=0$, whose
solution is~(\ref{m7}), again, for generic connections.
The Abelian gauge fixing (b) in~(\ref{m6}), and $A_{\mu}^A(x+a_\nu e_\nu)=
2\pi \partial_\mu \omega_{\nu}^A(x) + A^A_\mu (x)$ imply
\beq
\partial_\mu\partial_\mu \omega^A_{\nu}(x)=
[\partial_{\mu}A_{\mu}^A(x+a_{\nu}e_{\nu})-\partial_{\mu}A_{\mu}^A(x)]/2\pi
=0, \qquad \mbox{no sum on }\nu.
\eeq{m8}
On the $\omega^A_\mu(x)$, the consistency conditions~(\ref{OOOO}) become:
\beq
\omega_\mu^A(x+a_\nu e_\nu) + \omega_\nu^A(x) =
\omega_\nu^A(x+a_\mu e_\mu) + \omega_\mu^A(x)+ \hat{w}^A_{\mu\nu},\;\;
\mbox{no sum on }\mu,\;\nu.
\eeq{m9}
Notice that the definition of $\hw_{\mu\nu}$, given in eq.~(\ref{zmn}),
is not unique. One can always add to $\hw_{\mu\nu}\in \hL_W$ a root
$\hat{r}_{\mu\nu}\in \hL_R$.
This fact will play an essential role later on.
The general solution of eq.~(\ref{m9}) is
\beq
\omega_{\mu}(x)= \sum_\nu \left[ \tilde{w}_{\mu\nu} {x_\nu\over a_\nu} +
s_{\mu\nu}{x_\nu\over a_\nu} \right] + \tilde{\omega}_\mu(x).
\eeq{m10'}
{}From now on we shall drop the Cartan index $A$ whenever unambiguous; in this
equation
\ber
\tilde{w}_{\mu\nu}=\hw_{\mu\nu}, \qquad {\rm for} \,\,\,\,\,  \mu > \nu,
\nonumber\\
\tilde{w}_{\mu\nu}=0 \qquad {\rm otherwise},
\eer{m10''}
$s_{\mu\nu}^A$ are symmetric coefficients
in $\mu$, $\nu$, and $\tilde{\omega}^A_\mu(x)$ is a periodic function.
Applying eq.~(\ref{m8}) to (\ref{m10''}) gives
$\tilde{\omega}_\mu(x)=c_{\mu}$,  where $c_{\mu}$ is a constant and,
therefore,
\beq
\omega_{\mu}(x)= \sum_\nu \left[ \tilde{w}_{\mu\nu} {x_\nu\over a_\nu} +
s_{\mu\nu}{x_\nu\over a_\nu} \right] +  c_{\mu}.
\eeq{m10}
The boundary condition for the gauge field $A^A_\mu(x)$ becomes
\beq
A^A_\mu(x+a_\nu e_\nu)=A_{\mu}^A(x)+2\pi\partial_{\mu}\omega_{\nu}^A(x)=
A^A_\mu(x)+{2\pi\over a_{\mu}}\tilde{w}_{\nu\mu}^A +
{2\pi\over \a_{\mu}}s_{\nu\mu}^A,
\;\;\; \mbox{no sum on }\nu ,
\eeq{m11}
whose solution reads
\beq
A_\mu (x) = 2\pi \sum_\nu \left( \tilde{w}_{\nu\mu} {x_\nu\over a_\mu a_\nu}
+ s_{\mu\nu} {x_\nu\over a_\mu a_\nu} \right) + \tilde{A}_\mu(x).
\eeq{m12}
where $\tilde{A}_\mu(x)$ is periodic and $\sum_\mu s_{\mu\mu}/a_\mu^2=0$.

We can get rid of $s_{\mu\nu}$ exploiting the residual
invariance of the gauge-fixed
functional integral~(\ref{m5}) under a finite dimensional group of
transformations. Any gauge transformation $\O$ in the Cartan torus such that
\beq
\O (x+a_{\mu}e_{\mu})\O_{\omega_{\mu}}(x)\O^{-1}(x)=\O_{\omega'_{\mu}}(x),
\eeq{m13'}
where
\beq
\O_{\omega_{\mu}}(x)=e^{2\pi i \omega_{\mu}^A(x)T_A},
\eeq{m13''}
and both $\omega_{\mu}$ and $\omega'_{\mu}$ are of the form (\ref{m10}), is
a symmetry surviving the gauge fixing.
These gauge transformations, $\O(x)=\exp [2\pi i\omega^A(x)T_A]$, are
generated by
\beq
\omega^A (x) = \beta^A_{\mu\nu} x^\mu x^\nu + \gamma_{\mu}^A x^{\mu} +
\delta^A,
\eeq{m13}
with $\beta^A_{\mu\nu}$ a constant, symmetric, traceless matrix,
$\gamma_{\mu}^A$ a constant vector, and $\delta^A$ another arbitrary constant.
Equation (\ref{m13'}) is satisfied with
\ber
\omega'_{\mu}(x)&=&\omega_{\mu}(x)+\b_{\mu\mu}a_{\mu}^2
+\sum_{\nu}2\b_{\mu\nu}a_{\mu}a_{\nu}{x_{\nu}\over
a_{\nu}}+\gamma_{\mu}a_{\mu} \nonumber\\ &=&
\sum_\nu \left[ \tilde{w}_{\mu\nu} {x_\nu\over a_\nu} +
s'_{\mu\nu}{x_\nu\over a_\nu} \right] +  c'_{\mu}.
\eer{m14'}
Here
\beq
s'_{\mu\nu}=s_{\mu\nu}+2\b_{\mu\nu}a_{\mu}a_{\nu},\qquad
c'_{\mu}=c_{\mu}+\b_{\mu\mu}a_{\mu}^2+\gamma_{\mu}a_{\mu}.
\eeq{m14''}
By choosing $\beta^A_{\mu\nu}=-s^A_{\mu\nu}/2a_\mu a_\nu $,
$\gamma_{\mu}^A=-(c_{\mu}^A/a_{\mu}+\b_{\mu\mu}^A a_{\mu})$
we can fix both
$s_{\mu\nu}^A$ and $c^A_\mu$ to zero.

Now, eq.~(\ref{m5})
becomes
\ber
W[\vhk,\vhm]&=& {\cal V} \sum_{\hw_{\mu\nu}} \int
[d\tilde{A}^A_\mu(x)][dA^\alpha_\mu(x)] [d\Phi'(x)][dc(x)][d\bar{c}(x)]
\delta[\partial_\mu
\tilde{A}^A_\mu(x)] \delta[D_\mu^{Ab.}A_\mu^\alpha(x)]\nonumber \\
& & \exp\left\{-\int d^4x [L + \bar{c}(x) D_\mu^{Ab.}D_\mu c(x)]\right\}.
\eer{m14}
Here we have exponentiated the Jacobian $\delta F[\P]/\delta\Omega $ by
means of the standard Faddeev-Popov ghosts $c$, $\bar{c}$.
All non-gauge fields (scalars and spinors) are called $\Phi'(x)$, and the
sum over $\hw_{\mu\nu}$ is extended to all vectors in $\hL_W$ compatible
with~(\ref{OOOO}). If $\stackrel{o}{w}_{\mu\nu}$ is one such vector then,
as noticed after eq.~(\ref{m9}), the general $\hw_{\mu\nu}$ reads
\beq
\hw_{\mu\nu}=\stackrel{o}{w}_{\mu\nu}+\hat{r}_{\mu\nu}, \qquad
\stackrel{o}{w}_{\mu\nu}\in \hL_W/\hL_R, \,\,\,\,\, \hat{r}_{\mu\nu}\in \hL_R.
\eeq{m9'}
Thus the sum in eq. (\ref{m14}) runs over all elements $\hat{r}_{\mu\nu}$
of the dual root lattice $\hL_R$, translated by a  fixed dual weight
$\stackrel{o}{w}_{\mu\nu}$. We recall that the relation between
$\stackrel{o}{w}_{\mu\nu}$ and $\vhm,\vhk$ in $W[\vhk,\vhm]$ (\ref{m14}) is
given in eqs. (\ref{m}), (\ref{k}) with $\hw_{\mu\nu}$ being replaced by
$\stackrel{o}{w}_{\mu\nu}$.

The Abelian field strength
$F^A_{\mu\nu}=\partial_{\mu}A^A_{\nu}-\partial_{\nu}A^A_{\mu}$ reads
\beq
F^A_{\mu\nu}={2\pi\over a_{\mu}a_{\nu}} \hw^A_{\mu\nu} + \partial_{\mu}
\tilde{A}^A_{\nu}-\partial_{\nu}\tilde{A}^A_{\mu}.
\eeq{m15}
The boundary conditions of the superpartners of $A^A_{\mu}$, i.e. the
scalar field $\phi_{IJ}^A$ and the gauginos $\lambda_{I\,\alpha}^A$, are
given in eq.~(\ref{PO}). Explicitly, they read
\beq
\phi_{IJ}^A(x+a_{\mu}e_{\mu})=\phi_{IJ}^A(x), \;\;\; \lambda_{I\,\alpha}^A(x +
a_{\mu}e_{\mu})=-\lambda_{I\,\alpha}^A(x).
\eeq{m16}
The reason why we have chosen antiperiodic boundary conditions for the
fermions along all four directions, in eq.~(\ref{PO}) and here, is
that we want that the functional integrals $W[\vhk,\vhm]$ transform
into themselves under discrete $90^0$ Euclidean rotations. An example of
such rotations, giving rise to 't Hooft duality, is given in Section 5, in
equations~(\ref{so4}), (\ref{W1234}). Obviously the
quantities $w[\vhk,\vhm]$, defined in eq.~(\ref{m1}), transform under $O(4)$
exactly as $W[\vhk,\vhm]$.
Notice that if we had chosen periodic boundary conditions for the fermions, we
would have found a topological invariant, the Witten index~\cite{WI}. This
invariant does not depend on the coupling constant nor on theta angle,
and thus it is trivially S-duality invariant.

Equations~(\ref{m15}), (\ref{m16}) show that the twist affects only the vector,
among the fields of the Cartan supermultiplet. All fields with gauge indices
outside of the Cartan subalgebra obey {\em homogeneous} boundary conditions:
\beq
\Phi^{\alpha}(x +a_{\mu}e_{\mu})= (-)^F e^{2\pi i (\omega_{\mu}(x)\cdot
\alpha)}\Phi^\alpha(x),\;\;\; \alpha\in \Lambda_R.
\eeq{m17}
Notice that $\Phi^\alpha(x)=0$ satisfies these boundary conditions.

To compute the functional integral~(\ref{m14}) one must introduce an
ultraviolet regulator, for instance a momentum cutoff $\L_0$,
as well as the infrared regulator $M\beta V$, discussed at the beginning
of this Section\footnotemark.
\footnotetext{The cutoff $\L_0$ may break explicitly gauge invariance. In
any non-anomalous gauge theory this breaking can be reabsorbed by adding to
the bare action finite non-gauge invariant counterterms. The renormalization
approach used here is regularization-independent.}
We already mentioned
that in a finite box the infrared divergences are due to the existence of
flat directions in the scalar potential. These cannot be
lifted either perturbatively or non-perturbatively~\cite{seiberg}.
Let us label
these flat directions, which can be aligned along the Cartan subalgebra,
with $v_{IJ}^A$, and introduce the following notations:
\beq
U=\max_{I,J,A}|v_{IJ}^A|,\qquad  u=\min_{I,J,A}|v_{IJ}^A|.
\eeq{m17'}
Then the functional integral~(\ref{m14}) is {\em defined} as
\ber
W[\vhk,\vhm]&=& {\cal V} \lim_{M\rightarrow \infty} \lim_{\L_0
\rightarrow \infty} \int_{U\leq M}\prod_{I,J,A} dv_{IJ}^A
W_{v^A_{IJ}}[\L_0,\vhk,\vhm], \label{m18} \\
W_{v^A_{IJ}}[\L_0,\vhk,\vhm] &=& \sum_{\hw_{\mu\nu}} \int
[d\tilde{A}^A_\mu(x)][dA^\alpha_\mu(x)]
[d\hat{\phi}_{IJ}(x)][d\lambda_I(x)][dc(x)][d\bar{c}(x)]
\nonumber \\
& & \delta[\partial_\mu \tilde{A}^A_\mu(x)] \delta[D_\mu^{Ab.}A_\mu^\alpha(x)]
\exp\left\{-\int d^4x [L_{\L_0} + \bar{c}(x) D_\mu^{Ab.}D_\mu c(x)]\right\},
\nonumber \\
\eer{m19}
where, $\int d^4x\hat{\phi}^A_{IJ}(x)=\beta V v^A_{IJ}$, $\beta=a_4$,
$V=a_1a_2a_3$. The Lagrangian
$L_{\L_0}$ is regulated in an appropriate manner. A convenient
regularization scheme for our purposes is the Wilson-Polchinski
one~\cite{Wil,P,G,becchi}.
In this approach, one directly integrates the high-frequency
modes in the functional integral, and one defines the renormalization flow
by imposing the invariance of the functional integral as the ultraviolet
cutoff is changed. This allows us to compute the functional
integral~(\ref{m14}) in two steps, first integrating out the high-energy
modes, and then the low-energy ones.

More precisely, one modifies the inverse propagator by cutting out the
high-frequency modes. For a scalar of mass $m$, this is accomplished by
introducing a smooth momentum cutoff by the substitution~\cite{P}
\ber
& & p^2+ m^2 \rightarrow K^{-1}(p^2/\L^2)(p^2 +m^2),\nonumber\\
& & K(p^2/\L^2)=1,\;p^2\leq
\L^2-\epsilon,\;\;K(p^2/\L^2)=0,\; p^2\geq \L^2 +\epsilon,\;\; \epsilon >0.
\eer{m20}
This substitution eliminates all modes of frequency larger than $\sqrt{\L^2
+ \epsilon}$. The regulated $N=4$ Lagrangian is then defined as follows:
$L_{\L_0}=L_{kin\,\L_0} + L_{int\,\L_0}$. The interaction Lagrangian
$L_{int\,\L_0}$ contains all terms of order three or higher in the fields.
The kinetic action,
${\cal S}_{kin}=\int d^4x L_{kin}$
is quadratic in the fields and reads, in the momentum representation,
\beq
{\cal S}_{kin\, \L_0}= {1\over 4\pi}\mbox{Re}\, S \int {d^4p\over (2\pi)^4}
\P^*(p) \Pi(p)K^{-1}(p^2/\L_0)\P(p).
\eeq{m21}
$\P(p)$ denotes all the $N=4$ fields, as well as the gauge-fixing ghosts
in the momentum representation, whereas their unregulated kinetic term
is denoted by $\Pi(p)$. By varying the cutoff from $\L_0$ to $\L<\L_0$, one
``integrates out'' the modes of frequency $\L< \omega < \L_0$, to obtain an
effective action at the lower scale $\L$. This is the {\em Wilsonian}
effective action
\beq
{\cal S}_{\L}= {1\over 4\pi}\mbox{Re}\, S \int {d^4p\over (2\pi)^4}
\P^*(p) \Pi(p)K^{-1}(p^2/\L)\P(p) + \int d^4x L_{int\,\L}.
\eeq{m22}
The interaction Lagrangian at scale $\L$ depends on
$L_{int\,\L_0}$; it may contain a constant term independent of the fields,
and it is determined by imposing the equation
\beq
\L{\partial\over \partial\L}W_{v^A_{IJ}}[\L,\vhk,\vhm]=0,
\eeq{m23}
where
\ber
W_{v^A_{IJ}}[\L,\vhk,\vhm]&=&\sum_{\hw_{\mu\nu}} \int
[d\tilde{A}^A_\mu(x)][dA^\alpha_\mu(x)]
[d\hat{\phi}_{IJ}(x)][d\lambda_I(x)][dc(x)][d\bar{c}(x)]
\nonumber \\
&& \delta[\partial_\mu
\tilde{A}^A_\mu(x)] \delta[D_\mu^{Ab.}A_\mu^\alpha(x)]\exp(-{\cal S}_{\L}).
\eer{m24}
By definition $L_{int\,\L_0}$ is the bare interaction Lagrangian~\cite{P},
given in eq.~(\ref{L}).
Equation~(\ref{m24}) means that the functional integral~(\ref{m19}) can be
computed using the low-energy effective action $S_{\L}$, which, by
construction, contains only the degrees of freedom of frequency less than
$\L$.
The effective interaction Lagrangian $L_\L$ at the scale $\L$ contains
relevant operators (of dimension three and four) and irrelevant ones, of
dimension higher than four. By fixing the coefficients of these relevant
operators, (i.e. the renormalized coupling constants),
one defines implicitly the bare coupling constants at scale
$\L_0$. Renormalizability means that the limit $\L_0\rightarrow \infty$,
holding fixed the renormalized coupling constants,
exists and is unique. Polchinski~\cite{P} proved that this definition of
renormalizability coincides with the usual one. Thanks to the finiteness of
$N=4$,\footnotemark
\footnotetext{See \cite{White} for a scheme-independent proof and a review
of the literature about the problem.} a stronger statement holds in our case:
the limit
$\L_0\rightarrow\infty$ exists at fixed $S=4\pi/g^2 + i\theta/2\pi$, i.e. at
fixed {\em bare} coupling constant.

The advantage of eq.~(\ref{m24})
over eq.~(\ref{m19}) is that ${\cal S}_{\L}$
is much simpler than the original $N=4$ action. Indeed, ${\cal S}_{\L}$ can be
expanded in local operators when $v_{IJ}^A\neq 0$, and it reads
\beq
{\cal S}_{\L}=\int d^4 x \{ C[S(\L),\hw_{\mu\nu}] +
L_{N=4}[S(\L)] + \L^{-1} L^{(5)} +...\}.
\eeq{m25}
Here $L_{N=4}[S(\L)]$ is the dimension $\leq$ 4 Lagrangian. It is uniquely
fixed by $N=4$ supersymmetry. The only freedom is the choice of the effective
coupling constant and theta angle at the scale $\L$:
$S(\L)=4\pi/g^2(\L)+i\theta(\L)/2\pi$. The higher-dimension terms ($L^{(5)}$
etc.) are fixed by the renormalization conditions, or equivalently by the
bare coupling constants, thanks to the finiteness of $N=4$.
The field-independent constant $C[S(\L),\hw_{\mu\nu}]$ may depend in
principle on the
fluxes $\hw_{\mu\nu}$.\footnotemark
\footnotetext{Recall that our boundary conditions~(\ref{PO}) explicitly
break supersymmetry.}

A major simplification occurs if one chooses $\L\ll u$, where $u$ is
defined in eq. (\ref{m17'}). In this case one may
use the fact that all fields outside the Cartan subalgebra have masses
$O(gu)$, and use the decoupling theorem~\cite{AC}. This theorem was proven
in the contest of the Wilson-Polchinski scheme in~\cite{GZ}, and it states
that the effective action at $\L\ll u$ is
\beq
{\cal S}_{\L}= \int \{ C[S^{eff},\hw_{\mu\nu}] + L_{N=4}^{Ab.}[S^{eff}]
 \} + O(u^{-\varepsilon})
,\;\;\; \varepsilon > 0.
\eeq{m26}
The Lagrangian $L_{N=4}^{Ab.}$ is the (gauge fixed) $N=4$ one,
with all fields outside the Cartan subalgebra set to zero; thus, it is free
and quadratic, since an Abelian $N=4$ theory is noninteracting.
Equation~(\ref{m26})
gives a precise meaning to the physical intuition that very
heavy fields do not contribute to the long range (i.e. low energy) dynamics.

The constant $C[S^{eff},\hw_{\mu\nu}]$ arises from two sources. The
first is the shift in
the vacuum energy due to integrations over the high-energy modes. The
flux-dependent part of this contribution comes from the charged modes,
outside the Cartan subalgebra
(see eq. (\ref{m17})), since the neutral modes $\tilde{A}_\mu^A$,
$\hat{\phi}^A_{IJ}$, $\lambda^A_{I\,\alpha}$ in the Cartan
subalgebra do not depend at all on the fluxes $\hw_{\mu\nu}$. The
charged modes are massive $m\geq g^{eff}u\equiv
(4\pi\mbox{Re}\, S^{eff})^{-1/2}u$, therefore, they produce contributions to
the vacuum energy, whose
boundary-dependent part vanishes as
$mV^{1/3}\rightarrow\infty$ ($V^{1/3}$=size of the box).
Moreover,
to compute the functional integral, one must expand around {\em all}
local bosonic minima of the action (i.e. solutions of the classical
equations of motion with finite action),
and sum over all these minima. The expansion begins with a
field-independent term: the action computed on the classical
configuration, which obviously contributes to $C[S^{eff},\hw_{\mu\nu}]$.
Actually $L_{N=4}^{Ab.}$
already takes into account the contribution of purely Abelian
configurations. If we integrate over {\em all} $\O(x)$ obeying the consistency
condition~(\ref{OOOO}) any non-Abelian configuration necessarily involves
(non-gauge)
excitations of physical massive modes, since otherwise one could find a
gauge transformation bringing it back to an Abelian form.
To see whether non-Abelian configurations contribute to the
functional integral let us reason as follows.
The classical action of $N=4$ is conformally invariant. Thus we may rescale
coordinates and fields and express the bosonic action in terms of dimensionless
variables $x_{\mu}=y_{\mu}/u$,
$\phi_{IJ}^a(x) \rightarrow u^{-1}\phi_{IJ}^a(y)$,
$A_{\mu}^a(x) \rightarrow u^{-1}A_{\mu}^a(y)$. By construction, now
$\min_{I,J,A}|v^A_{IJ}|=1$ and
\beq
\int_{a_\mu}d^4 x L_{N=4}[u,\phi_{IJ}^a(x),A^a_{\mu}(x)]=\int_{ua_\mu}d^4y
L_{N=4}[1,\phi_{IJ}^a(y),A^a_{\mu}(y)].
\eeq{mm1}
The box, in the $y$ variables, has sizes $ua_\mu$. The classical action
${\cal S}_{N=4}$ contributes to the functional integral terms
$O[\exp(-{\cal S}_{N=4})]$.
We want to see what happens in the limit $u\rightarrow
\infty$, since our aim is to compute the leading term in the
expansion~(\ref{m1}). Thus, the only non-zero contributions are those whose
action is finite in the infinite-volume limit $ua_\mu\rightarrow \infty$.
The $N=4$ Lagrangian is given in eq.~(\ref{L}). It is immediate to see that
under the rescaling $\phi_{IJ}^a(y)\rightarrow \phi_{IJ}^a(\lambda
y)$,\footnotemark
\footnotetext{Notice that this rescaling does not change the constant mode
of the scalars.}
$A_{\mu}^a(y)\rightarrow \lambda^{-1} A_{\mu}^a(\lambda y)$, the bosonic
infinite-volume action scales as
\ber
{\cal S}_{N=4} & \rightarrow & \lambda^{-4} {\mbox{Re}\, S \over 4\pi}
\int d^4 y V[\phi_{IJ}^a(y)] +
\lambda^{-2} {\mbox{Re}\, S \over 4\pi}
\int d^4 y D_{\mu}\phi^{a\, IJ}(y)D^{\mu}\phi^a_{IJ}(y) + \nonumber \\
& & {{\rm Re}S\over 8\pi}\int d^4yF_{\mu\nu}^a(y)F^{a\,\mu\nu}(y) -
{i {\rm Im}S\over 8\pi}\int d^4yF_{\mu\nu}^a(y)\tilde{F}^{a\,\mu\nu}(y),
\nonumber \\
V[\phi_{IJ}^a(y)] & = &f_{abc}f_{ade}\phi^b_{IJ}(y)\phi^{c\,JK}(y)
\phi^d_{KL}(y) \phi^{e\,LI} (y) \geq 0
\eer{mm2}
Demanding as usual that $\lambda=1$ is a stationary point with
finite action, we find that both the potential and the kinetic term of the
scalars have to vanish, i.e. that the scalar fields are covariantly constant
and mutually commuting
\beq
D_{\mu}\phi^a_{IJ}(y)=0,\;\;\; f_{abc}\phi_{IJ}^b(y)\phi^{c\, JK}(y)=0.
\eeq{mm3}
This equation implies that the gauge connection is reducible, that is, it
belongs to the subgroup of
$G$ leaving $\phi_{IJ}^a$ invariant (i.e. the stabilizer of
$\phi_{IJ}^a(y)$).
With a gauge transformation, one can
reduce mutually commuting, covariantly constant scalars  to
$\phi_{IJ}(y)=v_{IJ}$. For
almost all $v_{IJ}^A$ the corresponding stabilizer group is the Cartan torus,
and the only gauge connections obeying~(\ref{mm3}) are the Abelian ones,
which already appear in the Wilsonian effective action~(\ref{m26}).

To sum up, the contribution of the constant
$C[S^{eff},\hw_{\mu\nu}]$ to the leading infrared divergence in
eq.~(\ref{m1}) does not depend on $\hw_{\mu\nu}$.

It is important to notice that if we choose to fix a particular set of
transition functions, $\stackrel{o}{\omega}_{\mu}(x)$ ($=\sum_{\mu\geq
\nu}\stackrel{o}{w}_{\mu\nu}x^\nu/a_\mu a_\nu$, for instance) and
integrate only on gauge transformations $\O(x)$ which leave invariant these
transition functions, we cannot reduce the computation of the functional
integral to the Abelian case. Indeed, we may
find non-Abelian classical configurations whose action stays finite in
the $u\rightarrow \infty$ limit, and which cannot be gauge-transformed into
Abelian configurations. Choose for instance periodic boundary
conditions: $\stackrel{o}{\omega}_{\mu}(x)=0$. Then
the configuration
\beq
A_{\mu}^A(x)=2\pi \sum_\nu \tilde{w}^A_{\mu\nu}{x_\nu\over a_\mu a_\nu},\;\;\;
\tilde{w}_{\mu\nu}^A \in \hL_R,\;\; \tilde{w}_{\mu\nu}^A \neq 0,
\eeq{mex}
(all other fields $=0$), has finite action
${\cal S}=\pi[\mbox{Re}\, S \beta V(\hw_{\mu\nu}\cdot \hw_{\mu\nu})/2a^2_\mu
a^2_\nu -i \mbox{Im}\, S \epsilon^{\mu\nu\rho\sigma}(\hw_{\mu\nu}\cdot
\hw_{\rho\sigma})/4]$.
Equation~(\ref{mex}) does not satisfy the boundary conditions
$\stackrel{o}{\omega}_{\mu}(x)=0$, but it can be changed into a
configuration which does it by
a {\em non-periodic} gauge transformation.
This gauge transformation is
non-Abelian, i.e. it involves fields outside of the Cartan subalgebra.
No {\em periodic} $\O(x)$ exists that brings back the gauge transformed
configuration into the Cartan subalgebra. In other words, the scaling
argument given above still works, but the scalars do not have a well
defined limit at infinity. The gauge transformation needed to bring the
reducible connection into the Cartan subalgebra does not belong to the
subspace of admissible transformations.
By summing over all gauge transformations, i.e. over $r_{\mu\nu}\in \hL_R$,
we enlarge the space of allowed transformations and thus we {\em can}
bring all reducible connections to the Cartan subalgebra.

The extra terms in eq.~(\ref{m26}) vanish in the limit $u\rightarrow
\infty$. The complexified, effective coupling constant $S^{eff}$ is a function
of the bare coupling constant, and, in general, it may depend on the VEVs
$v^A_{IJ}$, as well as on $\L$ and the renormalization scale $\mu$:
$S^{eff}=f(\L,v^A_{IJ},\mu,S)$.
Here, since the low-energy theory is free, $S^{eff}$ is independent of $\L$
(recall that $\L\ll u\equiv \min_{IJA}|v^A_{IJ}|$).
By dimensional reasons it can only depend on
dimensionless ratios
$S^{eff}=f(v^A_{IJ}/\mu,S)$. Since the beta function of $N=4$ is
equal to zero, $S^{eff}$ is independent of $\mu$, and thus of $v^A_{IJ}$:
$S^{eff}=f(S)$. This last equation defines $S^{eff}$ in terms of $S$, and
vice-versa\footnotemark, thus we may define our $N=4$ theory using $S^{eff}$
\footnotetext{$f(S)$ is invertible, at least in perturbation theory.}
instead of $S$. This we shall do, and from now on drop the superscript in
$S^{eff}$. In other words, from now on the coupling constant and theta
angle will denote the effective, infrared ones.

Finally, once the ultraviolet regulator $\L_0$ is removed,
we may compute the functional integral~(\ref{m18}) as follows.
First we split the integration over $v^A_{IJ}$ in two parts
\beq
W[\vhk,\vhm]= {\cal V}
\lim_{M\rightarrow \infty}\left\{
\int_{U\leq M,\; u\geq Q}\prod_{I,J,A} dv_{IJ}^A W_{v^A_{IJ}}[\vhk,\vhm]
+ \int_{U\leq M,\; u\leq Q}\prod_{I,J,A} dv_{IJ}^A W_{v^A_{IJ}}[\vhk,\vhm]
\right\}.
\eeq{m27}
Then we notice that $W[\vhk,\vhm]$ is divergent, but the coefficient
of the leading divergence, $w[\vhk,\vhm]$, is well defined.
It is also gauge invariant, and thus it is a good ``observable.''
Moreover, the integral over $u\leq Q$ does not contribute to the leading
infrared divergence, as long as $\lim_{M\rightarrow \infty}Q/M=0$. We may
choose for instance $Q=\log(M/\mu)$, with $\mu$ an arbitrary constant.
Now we may use eqs.~(\ref{m26}), (\ref{m24}) to get
\ber
W_{v^A_{IJ}}[\vhk,\vhm]&=& K[S]\sum_{\hw_{\mu\nu}} \int
[d\tilde{A}^A_\mu(x)][dA^\alpha_\mu(x)]
[d\hat{\phi}_{IJ}(x)][d\lambda_I(x)][dc(x)][d\bar{c}(x)]
\nonumber \\
&& \delta[\partial_\mu
\tilde{A}^A_\mu(x)] \delta[D_\mu^{Ab.}A_\mu^\alpha(x)]
\exp[-\int d^4x L_{N=4}^{Ab.}(S)][1+O(Q^{-\varepsilon})].
\eer{m28}
$w[\vhk,\vhm]$ is the $Q\rightarrow\infty$ limit of this equation.
The coefficient in front of~(\ref{m28}) may depend on the coupling constant but
is flux-independent.
Since the functional integral in eq.~(\ref{m28}) is quadratic,
it can be performed explicitly. The fluctuations about the
classical action modify the unknown constant $K[S]$, but
they do not introduce any dependence on $\hw_{\mu\nu}$.
Thus the overall scale in
front of $W[\vhk,\vhm]$ is flux-independent and
can always be redefined by shifting
the vacuum energy by a flux-independent constant.
The only
nonzero term in the classical action is the field strength~(\ref{m15}) with
$\tilde{A}_\mu^A=0$, and one finds
\beq
w[\vhk,\vhm]=\sum_{\hat{w}_{\mu\nu}} K'[S]
\exp\left[-\pi\sum_{\mu\nu}\left(
\beta V{{\rm Re}\, S \over 2 }{(\hw_{\mu\nu}\cdot \hw_{\mu\nu})
\over a_\mu^2 a_\nu^2} -i{{\rm Im}\, S \over 4}
\epsilon^{\mu\nu\rho\sigma}(\hw_{\mu\nu}\cdot\hw_{\rho\sigma})
\right)\right].
\eeq{m29}
Since the normalization constant $K'[S]$ is a matter of convention, we
may choose it, in particular, to be:
\beq
K'[S]=\left( {V(\mbox{Re}\, S)^3\over \b^3}\right)^{r/2},\;\;\;
r=\mbox{rank}\, G.
\eeq{mconst}

Equation~(\ref{m29}) gives us another test of S-duality, since $w[\vhk,\vhm]$
transforms under it as $W[\vhk,\vhm]$. In Section 4.2 we shall
derive the free energy
from $w[\vhk,\vhm]$. Its properties
under S-duality are examined in Section 6, while factorization, Witten
phenomenon, and its
transformation laws under 't Hooft's duality are discussed in Section 5.

Before coming to that, the following topics need to be discussed.
\begin{itemize}
\item
Since the functional integral~(\ref{W}) is
integrated over all $v^A_{IJ}$, it corresponds to a mixed thermodynamical
phase: the pure phases, in our gauge~(\ref{m6}),
are labelled by the VEVs
$\langle \phi_{IJ}^A\rangle$.
Moreover, only large VEVs contribute to eq.~(\ref{m29}). In other words,
$w[\vhk,\vhm]$ is ``blind'' to the phase diagram of $N=4$. One cannot argue
from eq.~(\ref{m29}) whether at $\langle\phi_{IJ}^A\rangle=0$ the
theory is in a non-Abelian Coulomb phase. Conformal invariance at zero VEVs,
though,
together with the fact that $N=4$ is in a Coulomb phase for all non-zero
VEVs, makes this hypothesis very plausible.
\item
In writing the integral~(\ref{m27}) we did not take into account that some
VEVs $v^A_{IJ}$ are identified under a gauge transformation: the gauge
fixing we chose still allows for constant gauge transformations in the Weyl
group of Lie $G$. This is a discrete group of finite order, say $|W|$.
Thus, given a $W_{v^A_{IJ}}[\vhk,\vhm]$, we should have
divided it by $d(v^A_{IJ})$, that is by the order of the orbit of $v^A_{IJ}$
under the Weyl group, to avoid overcounting
its contribution to the functional integral.
This proviso is irrelevant for our purpose, which is to compute
$w[\vhk,\vhm]$, since for large $v^A_{IJ}$ almost all orbits have dimension
$|W|$. Thus we may simply absorb $|W|$ in the definition of the partition
function.
\item
Finally, we may comment on what happens in $N<4$. One may think that we
never used in a fundamental way the
$N=4$ supersymmetry. Indeed we needed it to
ensure the vanishing of the beta function. If $\beta\neq 0$, the
complexified coupling constant $S$ would depend on $v^A_{IJ}$ in a very
non-trivial way~\cite{SW,YF}; in other words, the scalar potential would
receive radiative corrections, both at one loop and at the non-perturbative
level.
The computation of the integral over
$v^A_{IJ}$ would be ill-defined, since $S$ has a logarithmic singularity at
$v^A_{IJ}\rightarrow \infty$.  Moreover,
the integration over VEVs would wash
out the contribution of the most interesting points in the VEV space, namely
those where $S$ is singular, and extra massless degrees of freedom
(monopoles and dyons) appear in the low-energy effective action.
In the $N=4$ case, there is no point in looking for a standard QFT
description of the theory (with asymptotic particle states etc.) at the
point $v_{IJ}^A=0$. Indeed,
at that singular point, there exist both
massless electrically
charged particles {\em and} massless monopoles. This means that no local QFT
can describe the whole theory. The same conclusion arises if one notices
that conformally invariant theories do not have well defined asymptotic
states. At $v_{IJ}=0$
one should rather describe the theory using 4-D
superconformal invariance~\cite{FF}.
On the other hand, $N=2,1$ theories at their singular points
can still be described by a local QFT by
adding the appropriate massless degrees of freedom.
\end{itemize}
\subsection{From the Twisted Functional Integral to the Free Energy}
We are now ready to write down the leading infrared-divergent contribution
to the free energy, which can be derived by using eq.~(\ref{F}). This
equation says that the free energy is
the discrete Fourier transform of the twisted functional integrals;
thus its leading infrared divergence is the transform of  $w[\vhk,\vhm]$,
given in eq. (\ref{m29}). Then we will perform a Poisson resummation in the
vectors $\hw_{4i}$ to express the free energy in a form verifying and
generalizing the suggestion of ref.~\cite{GGPZ} to any compact,
simple group.

{}From eqs. (\ref{m}), (\ref{k}) and (\ref{m9'}) we have
\beq
\hw_{ij}=\e_{ijk}(\hl_k+\hm_k), \qquad \hw_{4i}=\hn_i+\hk_i, \qquad
i,j,k=1,2,3, \qquad \hk_i,\hm_i\in \hL_W/\hL_R, \,\,\, \hl_i,\hn_i\in
\hL_R.
\eeq{klmn}
Inserting into eqs. (\ref{m29}), (\ref{mconst}) one finds
\beq
w[\vhk,\vhm]=\left( {V\over \a^3 \b^3}\right)^{r/2}
\sum_{\hl_i,\hn_i\in \hL_R}e^{-S[\hl_i+\hm_i,\hn_i+\hk_i]},
\eeq{aa1}
where
\beq
S[\hl_i+\hm_i,\hn_i+\hk_i]=\pi\sum_{i=1}^3\left(\a^{-1}\Big[\bbar_i
(\hl_i+\hm_i)^2+\bbar_i^{-1}(\hn_i+\hk_i)^2\Big]
-2ia(\hl_i+\hm_i)(\hn_i+\hk_i)\right),
\eeq{aa2}
(recall (\ref{Saa}) that $\mbox{Re}\, S=\a^{-1}$, $\mbox{Im}\, S=a$).
In eq. (\ref{aa2}) and from now on, we define
\beq
\bbar_i=\beta{a_i^2\over V}.
\eeq{bbar}

By using eq. (\ref{F}) we find that the leading infrared-divergent
contribution to the free energy is
\ber
e^{-\b F[\ve,\vhm]}&=&{1\over N^3}\sum_{\vhk\in (\hL_W/\hL_R)^3}
e^{2\pi i \ve\cdot \vhk} w[\vhk,\vhm] \nonumber\\
&=& {1\over N^3}\left( {V\over \a^3 \b^3}\right)^{r/2} \sum_{\vhl\in\hL_R^3}
e^{-\pi\a^{-1}\sum_{i}\bbar_i(\hl_i+\hm_i)^2}  \times
\nonumber\\ &\times&
\sum_{\vhk\in (\hL_W/\hL_R)^3} \sum_{\vhn\in \hL_R^3}
e^{2\pi i[\ve\cdot(\vhn+\vhk)+a(\vhl+\vhm)\cdot (\vhn+\vhk)]}
e^{-\pi\a^{-1} \sum_{i} \bbar_i^{-1}(\hn_i+\hk_i)^2}.
\eer{aa3}
Here we have inserted $\exp (2\pi i\ve\cdot\vhn)=1$ inside the sum; this
identity holds because $\hL_R$ is the lattice dual to $\L_W$, and
because $\ve\in \L_W^3$, $\vhn\in \hL_R^3$ (see Appendix A). Now, we rewrite
\beq
\hw_i=\hn_i+\hk_i\in \hL_W
\eeq{aa4}
(this is what we previously called $\hw_{4i}$; here we drop the label 4),
and since the elements inside the sum depend on $\hn_i$ and $\hk_i$ only
through their sum, $\hw_i$, we may write the
double sum over $\vhk\in (\hL_W/\hL_R)^3$ and $\vhn\in \hL_R^3$ as a sum over
$\vhw$. We get
\beq
e^{-\b F[\ve,\vhm]}={1\over N^3}
\left( {V\over \a^3 \b^3}\right)^{r/2} \sum_{\vhl\in\hL_R^3}
e^{-\pi\a^{-1} \sum_{i}\bbar_i(\hl_i+\hm_i)^2} f(\ve,\vhm,\vhl),
\eeq{aa5}
where
\beq
f(\ve,\vhm,\vhl)=\sum_{\vhw\in \hL_W^3}
e^{2\pi i[\ve+a(\vhl+\vhm)]\cdot \vhw}
e^{-\pi\a^{-1} \sum_{i} \bbar_i^{-1} \hw_i^2}.
\eeq{aa6}
Expanding $e_i, \hw_i, \hl_i, \hm_i$ in the basis $e_n^*, \he_n^*, \he_n,
\he_n^*$, respectively, (the basis we use are described in Appendix A):
\ber
e_i=E_i^n e_n^*, \qquad \hw_i=K_i^n \he_n^*,  \,\, & & \,\,
\hl_i=L_i^n \he_n,\qquad \hm_i=M_i^n \he_n^*, \nonumber\\
E_i^n,K_i^n,L_i^n,M_i^n &\in& Z,
\eer{aa7}
we get
\beq
f(\ve,\vhm,\vhl)=\sum_{K_i^n=-\infty}^{\infty}
e^{2\pi i[E_i^t\g^t+a(L_i^t+M_i^t\g)B]K_i}
e^{-\pi\a^{-1} \sum_{i} \bbar_i^{-1} K_i^t C^{-1} K_i}.
\eeq{aa8}
Here $\g,C,B$ are the $r\times r$ matrices:
\beq
\g_{nm}=\he_n^*\cdot e_m^*, \qquad
C_{nm}=e_n\cdot e_m, \qquad B_{nm}=\he_n\cdot\he_m^*={1\over
2}\he_n^2\delta_{nm}, \qquad C^{-1}=\g B.
\eeq{aa9}

It is now the time to perform a Poisson resummation on the integers $K_i$.
The Poisson resummation formula is:
\beq
\sum_{K\in Z^n}
e^{-(K+\tau)^t A (K+\tau)}=
{\pi^{n/2}\over \sqrt{\det A}}\sum_{K\in Z^n}
e^{-\pi^2 K^t A^{-1} K} e^{2\pi i K^t \tau}.
\eeq{aa10}
Here $A$ is an $n\times n$ matrix, $K$ and $\tau$ are $n$-vectors.
Using the formula (\ref{aa10}) in eq. (\ref{aa8}) one finds
\beq
f(\ve,\vhm,\vhl)=(\det C)^{3/2}
\left({\a^3 \b^3 \over V}\right)^{r/2}\sum_{K_i\in Z^r}
e^{-\pi\a\sum_{i}\bbar_i[K_i^t+E_i^t\g^t+a(L_i^t+M_i^t\g)B]C
[K_i+\g E_i+aB^t(L_i+\g^t M_i)]}.
\eeq{aa11}
Inserting eq. (\ref{aa11}) into eq. (\ref{aa5}) we get
\ber
\exp\{-\b
F[E,M,S,\G]\}&=&c\prod_{i=1}^3\sum_{K_i^n,L_i^n=-\infty}^{\infty}\exp\Big\{
\nonumber\\
& &-\pi\bbar_i(K_i^n+E^p_i\g_{np},L_i^n+M^p_i\g_{pn})M(S,\G)
\left(\begin{array}{c}
K_i^m+E_i^p\g_{mp} \\ L_i^m+M_i^p\g_{pm}\end{array}\right)\Big\},
\nonumber\\
\eer{FEM}
where
\beq
M(S,\G)=\a
\left(\begin{array}{cc} (e_n\cdot e_m) & (e_n\cdot \he_m)a \\
                        (\he_n\cdot e_m)a  & (\he_n\cdot\he_m)(\a^{-2}+a^2)
\end{array}\right),
\eeq{MSG}
the matrix $\g$ is defined in (\ref{aa9}) and
\beq
c={1\over N^3}(\det C)^{3/2}.
\eeq{aa12}
To write eqs. (\ref{FEM}), (\ref{MSG}) we have used the relations:
\beq
(BC)_{nm}=\he_n\cdot e_m, \qquad (CB^t)_{nm}=e_n\cdot \he_m, \qquad
(BCB^t)_{nm}=\hat{C}_{nm}=\he_n\cdot \he_m.
\eeq{aa13}

Finally, one finds
\beq
\exp\{-\b F[\ve,\vhm,S]\}=c\prod_{i=1}^3\sum_{k_i\in \L_R, \hl_i\in \hL_R}
\exp\Big\{
-\pi\bbar_i(k_i+e_i,\hl_i+\hm_i)M(S)\left(\begin{array}{c}
k_i+e_i \\ \hl_i+\hm_i\end{array}\right)\Big\},
\eeq{FS}
recall that $\bbar_i=\b a_i^2/V$ contain all the information about the
sizes of the spatial box, $a_i$, and the temperature $\b^{-1}$.
In eq. (\ref{FS}), the factor
$c$ is a constant, independent of the fluxes $\ve$ and $\vhm$
and independent of $S$, and $M(S)$ is a $2\times 2$ matrix
\beq
M(S)={1\over {\rm Re}\, S}
\left(\begin{array}{cc} 1 & {\rm Im}\, S \\  {\rm Im}\, S & |S|^2
\end{array}\right)=\a
\left(\begin{array}{cc} 1 & a \\  a & \a^{-2}+a^2 \end{array}\right).
\eeq{MS}

If $\G$ is simply-laced then $(e_n\cdot e_m)=C_{nm}$ is the
Cartan matrix of the Lie algebra and, moreover,
\ber
&& \he_n=e_n, \,\, \he_n^*=e_n^*  \implies
e_n\cdot e_m=\he_n\cdot e_m=e_n\cdot \he_m=\he_n\cdot \he_m=C_{nm},
\nonumber \\ & & \g_{nm}=(C^{-1})_{nm}.
\eer{eeg}
Therefore,
\ber
&&\exp\{-\b F[E,M,S,\;{\rm simply-laced}\;\G]\}
=c\prod_{i=1}^3\sum_{K_i^n,L_i^n=-\infty}^{\infty}\exp\Big\{
\nonumber\\
&&
-\pi\bbar_i\Big(K_i^n+(E_iC^{-1})^n,L_i^n+(M_iC^{-1})^n\Big)C_{nm}\otimes M(S)
\left(\begin{array}{c}
K_i^m+(C^{-1}E_i)^m \\ L_i^m+(C^{-1}M_i)^m\end{array}\right)\Big\}.
\eer{FEMsl}
It is remarkable that eq. (\ref{FEMsl}) is formally equal to the classical
piece of a twisted
genus-1 string partition function on a toroidal background;
the genus-1 modular
parameter is $S$, the target-space background matrix is $C\otimes
I_{3\times 3}$, and the twist is $(E_i,M_i)$~\cite{GPR}.
\section{Properties of the Free Energy}
\setcounter{equation}{0}
In this Section we discuss some properties of the free energy:

\subsection{Factorization}
The free energy in eq. (\ref{FS}) factorizes at $\theta=0$:
\beq
F[\ve,\vhm,g,\theta=0]=F[\ve,0]+F[0,\vhm]+c,
\eeq{fac}
where $c$ is independent of the fluxes $\ve$ and $\vhm$.
Such a factorization is physically quite plausible if one assumes the
absence of interference between electric and
magnetic fluxes in the limit $a_i,\beta\to\infty$ \cite{tH}.
This holds not only for Abelian fields, as soon as
one assumes that strings occupy only a negligible portion of the total
space whereas magnetic fields fill the whole space.

An $N=4$ supersymmetric Yang-Mills theory is
scale invariant in an infinite box.
In Section 4 we have isolated the leading infrared-divergent
contribution to the free energy and, therefore,  we expect
the free energy in eq. (\ref{FS}) to be scale invariant. Indeed, it
is invariant under the scale transformation
\beq
F[La_i,L\b]=F[a_i,\b],
\eeq{sca}
thus, factorization in a large box implies factorization for any
size of the box. Indeed, factorizability at $\theta=0$ is obtained.

\subsection{Witten's Phenomenon}
The free energy for non-zero $\theta$ is derived from the free energy at
$\theta=0$ by the shift of $e_i$ and $k_i$
in the exponent inside the sum over $k_i$ and $\hl_i$ in eq. (\ref{FS}):
\beq
e_i\to e_i+{\theta\over 2\pi}\hm_i, \qquad
k_i\to k_i+{\theta\over 2\pi}\hl_i,
\eeq{witten}
Explicitly:
\ber
\exp\{-\b F[\ve,\vhm,S]\}&=&c\prod_{i=1}^3\sum_{k_i\in \L_R, \hl_i\in \hL_R}
\exp\Big\{ \nonumber\\ &&
-\pi\bbar_i(k_i+e_i+{\theta\over 2\pi}(\hl_i+\hm_i),\hl_i+\hm_i)M(g)
\left(\begin{array}{c}
k_i+e_i+{\theta\over 2\pi}(\hl_i+\hm_i) \\ \hl_i+\hm_i\end{array}\right)\Big\},
\nonumber \\
\eer{equa}
where
\beq
M(g)=M(S)|_{\theta=0}=
\left(\begin{array}{cc} {g^2\over 4\pi} & 0 \\  0 & {4\pi\over g^2}
\end{array}\right).
\eeq{Mg}
This is the Witten phenomenon \cite{W}.

To understand why both the non-Abelian electric fluxes, $e_i$, and the root
lattice vectors, $k_i$, are shifted when $\theta$ is turned on, let us discuss
the physical meaning of $k_i\in \L_R$.
The difference with the Abelian case is that the non-Abelian electric
and magnetic fluxes are defined modulo elements of the root lattices. The
physical reason for this indeterminacy is the following~\cite{tH}. The
theory contains elementary massive, charged particle carrying charge
$\alpha$ under the Cartan $U(1)^r$; here $\alpha$ is a root of $G$.
By pair-producing $n$ such
particles, and letting them wind around the box once in the direction $i$,
before annihilating them, we may change the electric flux along $i$ by
$k=n\alpha\in \L_R$. The change in energy due to this process is finite:
$g^2n^2(\alpha\cdot \alpha)/a_i$. The same argument applies to the magnetic
flux, $\hm_i$ which is changed by $\hl_i\in \hL_R$, {\em mutatis mutandis}.
Therefore, in a given flux sector, $k_i,\hl_i$,
the total electric flux, $e_i+k_i$, is shifted
by the total magnetic flux, $\hm_i+\hl_i$, when we turn on theta.

Witten's phenomenon also implies that when $\theta\to
\theta+2\pi$, the free energy of electric flux $\ve$ should transform into
the free  energy of electric flux $\ve+\vhm$. It is therefore important
that
\beq
e_i+\hm_i\in \L_W,\qquad k_i+\hl_i\in \L_R.
\eeq{cons}
Let us check that this is indeed the case. Expanding $e_i\in \L_W$ and
$\hm_i\in \hL_W$ in the basis (\ref{eLW}) and (\ref{heW}), respectively,
\beq
e_i=E_i^n e_n^*, \qquad \hm_i=M_i^n \he_n^*, \qquad E_i^n,M_i^n\in Z,
\nonumber\\
\eeq{1}
we get
\ber
e_i=E_i^n e_n^*&\to& e_i+\hm_i=E_i^n e_n^*+M_i^n \he_n^*\, \implies\nonumber \\
E_i^n e_n^* &\to& \sum_n(E_i^n+M_i^m \he_m^*\cdot \he_n)e_n^*
=\sum_n[E_i^n+{1\over 2}(\he_n\cdot \he_n)M_i^n]e_n^*.
\nonumber\\
\eer{2}
Here we used eqs. (\ref{hee}) and (\ref{heeeW}) of Appendix A.
Now, from eq. (\ref{he24})
we learn that $E_i^n+{1\over 2}(\he_n\cdot \he_n)M_i^n\in Z$ and,
therefore, $e_i+\hm_i\in \L_W$.
Similarly, one can show that $k_i+\hl_i\in \L_R$, and this completes the
consistency check for $\theta\to \theta+2\pi$.

\subsection{The 't Hooft Duality}
Clearly, the functional integral $W[\vhk,\vhm]$,
and the coefficient $w[\vhk,\vhm]$ of the leading
infrared divergence, must be invariant under joint
rotations of $a_{\mu}$ and $\hw_{\mu\nu}$ in Euclidean space. In particular,
if we perform the $SO(4)$ rotation corresponding to the interchange
$1\leftrightarrow 2$, $3\leftrightarrow 4$, represented by the matrix
\beq
\left(\begin{array}{cccc} 0&-1&0&0 \\ 1&0&0&0\\ 0&0&0&1 \\ 0&0&-1&0
\end{array}\right),
\eeq{so4}
and keeping in mind the relation between $\hm_i$, $\hk_i$ and $\hw_{\mu\nu}$
given in eqs. (\ref{m}), (\ref{k}), respectively, we find
\beq
w[\hk_1,\hk_2,\hk_3,\hm_1,\hm_2,\hm_3; a_1,a_2,a_3,\b]=
w[\hm_1,\hm_2,\hk_3,\hk_1,\hk_2,\hm_3; a_2,a_1,\b,a_3].
\eeq{W1234}
Indeed, $w[\vhk,\vhm]$ in eq. (\ref{m29}) manifestly obeys (\ref{W1234}).

The consequence of eq. (\ref{W1234}) for the free energy is
\ber
&&\exp\{-\b F[e_1,e_2,e_3,\hm_1,\hm_2,\hm_3; a_1,a_2,a_3,\b]\}=
{1\over N^2}\sum_{\hk_1,\hk_2\in \hL_W/\hL_R,\, l_1,l_2\in \L_W/\L_R}
\nonumber\\
&&\exp\{2\pi i(\hk_1\cdot e_1+\hk_2\cdot e_2-l_1\cdot\hm_1-l_2\cdot\hm_2)\}
\exp\{-a_3F[l_1,l_2,e_3,\hk_1,\hk_2,\hm_3,a_2,a_1,\b,a_3]\},
\nonumber \\
\eer{F1234}
where $N=Order(C)$.
This is the 't Hooft duality relation \cite{tH}.
Indeed, the free energy in (\ref{FEM}) obeys (\ref{F1234}). To prove it,
directly for the free energy, one should perform a Poisson resummation on
the directions $i=1,2$, as was done in \cite{GGPZ}. Obviously, here there is
nothing to prove since we have computed the functional integrals $w$ and,
therefore, 't Hooft's duality is automatic.

\section{S-Duality}
\setcounter{equation}{0}
The S-duality group is generated by the elements $\S,\T$, acting
on $S$ by:
\beq
\S \, : \qquad  S\to {1\over S},
\eeq{SS}
\beq
\T \, : \qquad  S\to S+i,
\eeq{TT}
or, as we will show soon, acting on $\G$, $\ve$ and $\vhm$ by:
\beq
\S \, : \qquad  \G\leftrightarrow\hg\,\,\,\, {\rm together}\,\,{\rm
with}\,\,\,\, \ve\to \vm, \,\, \vhm\to -\vhe ,
\eeq{SS1}
\beq
\T \, : \qquad \ve\to \ve+\vhm .
\eeq{TT1}
By the interchange of the Lie algebra with its dual algebra,
$\G\leftrightarrow\hg$ in (\ref{SS1}),
we mean: $\L_{R,W}\leftrightarrow\hL_{R,W}$, {\em i.e.},
$e_n\leftrightarrow\he_n$, $e_n^*\leftrightarrow\he_n^*$.
By $\ve\to \vm$, $\vhm\to -\vhe$ in (\ref{SS}) we mean: $E_i^n\to M_i^n$,
$M_i^n\to -E_i^n$, where the integers
$E_i^n,M_i^n$ are defined in eq. (\ref{1}).
The shift of $\ve$ by $\vhm$ in (\ref{TT1}) is consistent, because
$e_i+\hm_i\in \L_W$; this was checked in Section 5.2.

The elements $\S$ and $\T$ generate a group isomorphic to $SL(2,Z)$, acting
on $iS$ by the fractional linear transformations:
\beq
iS\to\left(\begin{array}{cc} a & b \\  c & d \end{array}\right) (iS)=
{a(iS)+b\over c(iS)+d},
\eeq{Sabcd}
where
\beq
\left(\begin{array}{cc} a & b \\  c & d \end{array}\right)\in
SL(2,Z),\qquad a,b,c,d\in Z, \qquad ad-bc=1.
\eeq{abcd}
The elements $\S$ and $\T$ correspond, therefore, to the matrices
\beq
\S=\left(\begin{array}{cc} 0 & -1 \\  1 & 0 \end{array}\right), \qquad
\T=\left(\begin{array}{cc} 1 & -1 \\  0 & 1 \end{array}\right).
\eeq{ST}
At $\theta=0$, the duality transformation
$\S$ takes $\a$ to $1/\a$ (namely, $\S\, : \, g\to 4\pi/g$), and it is
therefore called ``strong-weak coupling duality.'' Alternatively, it
interchanges the electric flux $\ve$ with the magnetic flux $\vhm$ and,
therefore, it is also called ``electric-magnetic duality.''
The transformation $\T$ simply shifts the theta parameter by $2\pi$:
$\T\, : \, \theta\to\theta+2\pi$.

When $\G$ is simply-laced, any element of $SL(2,Z)$ in (\ref{abcd})
transforms the electric and magnetic fluxes into other permitted fluxes. This
is not true, however, if $\G$ is non-simply-laced. Therefore, in the
following we shall
discuss the simply-laced and the non-simply-laced cases separately, when
required.

\subsection{The Action of S-Duality on the Free Energy for
Simply-Laced $\G$}
If $\G$ is simply-laced, the Lie algebra and its dual algebra are equal:
$\G=\hg$, namely, $\L_R=\hL_R$, $\L_W=\hL_W$. The free energy $F[E,M,S]$,
given in eq. (\ref{FEMsl}), transforms covariantly under $SL(2,Z)$
S-duality:
\beq
F\left[E,M,{1\over i}{a(iS)+b\over c(iS)+d}\right]=
F[dE-bM,aM-cE,S].
\eeq{SofF}
To prove eq. (\ref{SofF}) we note that S-transformation in eq. (\ref{Sabcd})
transforms $M(S)$ (\ref{MS}) as follows:
\beq
M\left({1\over i}{a(iS)+b\over c(iS)+d}\right)=AM(S)A^t, \qquad
A=\left(\begin{array}{cc} d & -c \\  -b & a \end{array}\right).
\eeq{Mabcd}
Let us first show how to get eq. (\ref{Mabcd}).
Let $g$ be an element of $SL(2,R)$, represented by a $2\times 2$ matrix,
and acting on a complex variable,
$x$, by a fractional linear transformation:
\ber
g=\left(\begin{array}{cc} a & b \\  c & d \end{array}\right), \qquad
a,b,c,d&\in& R, \qquad ad-bc=1, \nonumber\\
g(x)={ax+b\over cx+d}, \qquad x&\in& C .
\eer{gx}
Let $g_{S}$ be the element
\beq
g_S=\left(\begin{array}{cc} \a^{-1/2} & -a\a^{1/2} \\  0 & \a^{1/2}
\end{array}\right),
\eeq{gSaa}
where $S$, $\a$ and $a$ are given in eq. (\ref{Saa}).
One finds that
\beq
g_S(i)=i\a^{-1}-a=iS,
\eeq{gSi}
and, therefore,
\ber
g_{S'}(i)=iS'&\equiv& {a(iS)+b\over c(iS)+d}=g(iS)=g(g_S(i))=(gg_S)(i)
\nonumber \\
&\implies& g_{S'}=gg_S .
\eer{ggS}
Moreover, one finds that
\beq
M(S)=\e g_S g_S^t \e^t, \qquad
\e=\left(\begin{array}{cc} 0 & 1 \\  -1 & 0 \end{array}\right),
\eeq{Mgg}
where $M(S)$ is given in (\ref{MS}). Now, by using eqs. (\ref{Mgg}) and
(\ref{ggS}) we find
\beq
M(S')=\e g_{S'}g_{S'}^t\e^t=\e g g_S g_S^t g^t\e^t
=(\e g\e^t)(\e g_S g_S^t\e^t)(\e g^t\e^t)=AM(S)A^t,
\eeq{MS'}
where
\beq
A=\e g\e^t=\left(\begin{array}{cc} d & -c \\  -b & a \end{array}\right).
\eeq{Aege}
With the definition of $S'$ in eq. (\ref{ggS}),
this completes the proof of eq. (\ref{Mabcd}).
Finally, by using (\ref{Mabcd}) in eq. (\ref{FEMsl}) one can derive the result
(\ref{SofF}).

\subsection{The Action of the Generators on the Fluxes}
Equation (\ref{Mabcd}) can be used to prove (\ref{SS1}) and (\ref{TT1}).
The exponent of the free energy in eq. (\ref{FS}) transforms to
\ber
\exp\{-\b F[\ve,\vhm,S]\}&\to& \exp\{-\b F[\ve,\vhm,(a(iS)+b)/i(c(iS)+d)]\}
\nonumber\\
&=&c\prod_{i=1}^3\sum_{k_i\in \L_R, \hl_i\in \hL_R}
\exp\Big\{ \nonumber \\ &&
-\pi\bbar_i(k_i+e_i,\hl_i+\hm_i)AM(S)A^t\left(\begin{array}{c}
k_i+e_i \\ \hl_i+\hm_i\end{array}\right)\Big\},
\eer{FStrans}
where $A$ is given in eq. (\ref{Mabcd}).
For the transformation $\S$ in (\ref{ST}), $a=d=0, c=-b=1$:
\ber
\S &:& A=\left(\begin{array}{cc} 0 & -1 \\  1 & 0 \end{array}\right)
\implies \nonumber\\
\exp\{-\b F[\ve,\vhm,1/S]\}&=&c\prod_{i=1}^3\sum_{k_i\in \L_R, \hl_i\in \hL_R}
\exp\Big\{ -\pi\bbar_i(\hl_i+\hm_i,-k_i-e_i)M(S)\left(\begin{array}{c}
\hl_i+\hm_i\\ -k_i-e_i\end{array}\right)\Big\}
\nonumber\\
&=&c\prod_{i=1}^3\sum_{k'_i\in \L_R, \hl_i\in \hL_R}
\exp\Big\{ -\pi\bbar_i(\hl_i+\hm_i,k'_i-e_i)M(S)\left(\begin{array}{c}
\hl_i+\hm_i\\ k'_i-e_i\end{array}\right)\Big\}
\nonumber\\
&=&\exp\{-\b F[\vhm,-\ve,S]\}.
\eer{Str}
Here we have used the fact that $k_i\in \L_R\, \implies\, k'_i=-k_i\in \L_R$.
The result (\ref{Str}) proves eq. (\ref{SS1}).
For the transformation $\T$ in (\ref{ST}), $a=-b=d=1, c=0$:
\ber
\T &:& A=\left(\begin{array}{cc} 1 & 0 \\  1 & 1 \end{array}\right)
\implies \nonumber\\
\exp\{-\b F[\ve,\vhm,S+i]\}&=&c\prod_{i=1}^3\sum_{k_i\in \L_R, \hl_i\in \hL_R}
\exp\Big\{  \nonumber\\
& &-\pi\bbar_i(k_i+\hl_i+e_i+\hm_i,\hl_i+\hm_i)M(S)\left(\begin{array}{c}
k_i+\hl_i+e_i+\hm_i \\ \hl_i+\hm_i \end{array}\right)\Big\}
\nonumber\\
&=&c\prod_{i=1}^3\sum_{k'_i\in \L_R, \hl_i\in \hL_R}
\exp\Big\{  \nonumber\\
& &-\pi\bbar_i(k'_i+e_i+\hm_i,\hl_i+\hm_i)M(S)\left(\begin{array}{c}
k'_i+e_i+\hm_i\\ \hl_i+\hm_i \end{array}\right)\Big\}
\nonumber\\
&=&\exp\{-\b F[\ve+\vhm,\vhm,S]\}.
\eer{Ttr}
Here we have used the fact that $k_i\in \L_R,\,\, \hl_i\in \hL_R\, \implies\,
k_i+\hl_i\in \L_R$ (see Section 5.2).
The result (\ref{Ttr}) proves eq. (\ref{TT1}).

Finally, let us emphasize that eqs. (\ref{Str}) and (\ref{Ttr}) are correct
for any $\G$, either simply-laced or non-simply-laced.

\subsection{The Action of S-Duality on the Free Energy for
Non-Simply-Laced  $\G$}

{}From eqs. (\ref{Str}) and (\ref{Ttr}) we read that
\ber
\S(F[\ve,\vhm,S])&=&F[\vhm,-\ve,S], \nonumber \\
\T(F[\ve,\vhm,S])&=&F[\ve+\vhm,\vhm,S],
\eer{a0}
no matter whether
$\G$ is simply-laced or not. In particular, in both cases,
\beq
\S^2(F[\ve,\vhm,S])=F[-\ve,-\vhm,S].
\eeq{a1}
But there is a difference between the simply-laced case and the
non-simply-laced case.
If $\G$ is {\em non}-simply-laced, the Lie algebra and its dual algebra are
different:
$\G=so(2n+1)\Leftrightarrow\hg=sp(2n)$. In Appendix A it is shown that if the
long roots of $\hg$ are normalized with square
length 2 then the long roots of
the dual algebra have square length 4, namely, $\hL_R=\sqrt{2}\L_R(\hg)$,
$\hL_W=\sqrt{2}\L_W(\hg)$. Therefore, if in $\S(F)$ we renormalize the
roots such that long roots have square length 2, we obtain:
\beq
\S(F[\ve,\vhm,S])=F[\ve,\vhm,1/S]=F[\vhm,-\ve,S]
=F[\vhm/\sqrt{2},-\sqrt{2}\ve,S/2].
\eeq{a2}
In $F[\vhm/\sqrt{2},-\sqrt{2}\ve, S/2]$, the sum is over vectors
$\hk_i+\hm_i/\sqrt{2}$, where $\hk_i\in \L_R(\hg)$, and over vectors
$l_i-\sqrt{2}e_i$, where $l_i\in \hL_R(\hg)$ (recall that $\L_R(\hg)$ is
normalized such that roots have square length 2 or 1, and $\hL_R(\hg)$ is
normalized such that roots have square length 4 or 2).
But with this normalization $S$ is changed to $S/2$, and this is the
content of the last equality in eq. (\ref{a2}).

It might be helpful to rewrite the transformation of
the free energy in its presentation $F[E,M,S,\G]$,
given in eq. (\ref{FEM}); it transforms under $\S$ in (\ref{ST}) in the
following way:
\beq
F[E,M,1/S,\G_{e^2_{long}=2}]=F[M,-E,S,\hg_{e^2_{long}=4}]=
F[M,-E,S/2,\hg_{e^2_{long}=2}].
\eeq{SSofF}
Here $\G_{e^2_{long}=n}$ describes a Lie algebra $\G$ with long roots
normalized to have square length equal to $n$.

Equation (\ref{a2}) is by no means problematic. We do, however, run into
a problem if we follow an $\S$ transformation by $\T$.
In fact, acting on $F[\ve,\vhm,S]$ with $\T\S$ is not consistent. By using
eqs. (\ref{Str}) and (\ref{Ttr}) we get:
\beq
\T\S (F[\ve,\vhm,S])=\T (F[\vhm,-\ve,S])=F[\vhm-\ve,-\ve,S].
\eeq{a3}
But $\vhm-\ve$ is not a vector in $\hL_W^3$ ($\vhm-\ve \in \L_W^3$ (see
Section 5.2), but if $\G$
is non-simply-laced, $\L_W$ is not in $\hL_W$. Explicitly,
\beq
\hm_i-e_i=M_i^n\he_n^*-E_i^n e_n^*=\sum_n(M_i^n-E_i^m e_m^*\cdot
e_n)\he_n^*=\sum_n\Big[M_i^n-{1\over 2}(e_n\cdot e_n)E_i^n\Big]\he_n^*,
\eeq{a3'}
and since $e_n\cdot e_n=1$ or 2, $M_i^n-{1\over 2}(e_n\cdot e_n)E_i^n$ is
not necessarily an integer and, therefore, $\hm_i-e_i$ is not in $\hL_W$.
In other words, the
normalized flux, $(\vhm-\ve)/\sqrt{2}$ is not in $\L_W(\hg)$ or,
equivalently, the shift $S/2\to S/2+i/2$ in
$F[\vhm/\sqrt{2},-\sqrt{2}\ve,S/2]$ (\ref{a2}) is not a $2\pi$ theta shift
of the coupling $S/2$.)
Therefore, $\vhm-\ve$ is an ``illegal'' electric flux.
However, it is consistent to act on $F[\ve,\vhm,S]$ with $\T^2\S$, since we
get
\beq
\T^2\S (F[\ve,\vhm,S])=F[\vhm-2\ve,-\ve,S].
\eeq{a4}
Now $\vhm-2\ve\in \hL_W^3$, so it is a ``legal'' electric flux in the $\hg$
theory. (Equivalently, $S/2\to (S+2i)/2=S/2+i$ is a $2\pi$ theta shift of
the coupling $S/2$ in eq. (\ref{a2}).)

The transformation of $F[\ve,\vhm,S]$ under $\T$ in eq. (\ref{a0}) is
consistent, of course, because $\ve+\vhm\in\L_W$ is a legal flux.
It might be helpful to write the
transformation of $F[E,M,S,\G]$, given in eq. (\ref{FEM}),
under $\T$ in (\ref{ST}); it reads:
\beq
F[E^n,M^n,S+i,\G]=F\left[E^n+{1\over 2}(\he_n\cdot \he_n)M^n,M^n,S,\G\right].
\eeq{TofF}
Recall (\ref{he24}) that $\he_n\cdot \he_n=2$ or 4 and, therefore,
$E^n+(\he_n\cdot \he_n)M^n/2$ is an integer, as required.

To summarize, the elements of $SL(2,Z)$ which transform $F[\ve,\vhm,S,\G]$
to $F[\ve',\vhm',S,\G]$, where $\ve,\ve'\in \L_W^3$, $\vhm,\vhm'\in
\hL_W^3$, are
generated by $\T$ and $\S\T^2\S$. These elements generate a subgroup of
$SL(2,Z)$, called $\Gamma_0(2)$.
However, when we act on $F[\ve,\vhm,S]$ with an element of $SL(2,Z)$
containing  an odd number of the generator $\S$, the free energy for
the Lie algebra $\G$ is transformed
into a free energy for the dual Lie algebra $\hg$. In the simply-laced
case, $\hg=\G$, but in the non-simply laced case $\hg\neq \G$.
Moreover, in the latter case there are elements of $SL(2,Z)$ that transform
$F[\ve,\vhm,S]$ to a free energy of ``illegal'' fluxes; for example,
$\T\S$ is illegal.

\subsection{The Partition Function and Electric-Magnetic Duality}
So far the discussion about the free energy in a given flux sector,
$F[\ve,\vhm]$, did not require the use of the gauge {\em group}, but only
the properties of its Lie algebra. However, given a gauge group, $G$, not
all flux sectors are permitted \cite{GNO}; the electric fluxes, $\ve$,
are in the weight lattice of the group $G$ (modulo the root lattice of $G$):
\beq
e_i\in {\L_W(G)\over \L_R(G)}, \qquad i=1,2,3,
\eeq{einLW}
and the magnetic fluxes, $\vhm$, are in the weight lattice of the dual group
$\hG$ (modulo the root lattice of $\hG$, and up to a normalization if $\G$
is not simply-laced):
\beq
m_i\in {\hL_W(G)\over \hL_R(G)}, \qquad i=1,2,3.
\eeq{minLW}
(We recall that $\hL(G)_{R,W}=N(\G)\L(\hG)_{R,W}$,
where $N(\G)=1$ if $G$ is simply-laced, and
$N(\G)=\sqrt{2}$ if $G$ is non-simply-laced. In general,
$\L_W(G)$ is a sub-lattice of the weight
lattice of the Lie algebra $\L_W(\G)$, which is the weight lattice of the
universal covering group. The weight lattice of the dual group, $\L_W(\hG)$
is dual to the weight lattice of $G$. See Appendix A for more details.)

The partition function for a group $G$, $\Z(G)$, is given by summing over
the allowed flux sectors:
\beq
\Z(G,S)=\sum_{\ve\in (\L_W(G)/\L_R(G))^3,\,\, \vhm\in (\hL_W(G)/\hL_R(G))^3}
e^{-\b F[\ve,\vhm,S]}.
\eeq{ZGS}
Therefore, under strong-weak coupling duality, $\S$, the partition function
transforms to
\ber
\S &:& \,\, \Z(G,S)\to \Z(G,1/S)=\Z(\hG,S), \qquad\,\,\, {\rm if}\,\,\,\,\,
G \,\, {\rm simply-laced} \nonumber\\
\S &:& \,\, \Z(G,S)\to \Z(G,1/S)=\Z(\hG,S/2), \qquad {\rm if}\,\, G \,\, {\rm
non-simply-laced}.
\nonumber \\
\eer{SofZ}
This equation tells us that {\em the partition function of an $N=4$ Yang-Mills
theory with a coupling constant $1/S$ and a simply-laced (non-simply-laced)
gauge group $G$ is identical to the partition function of a theory with
coupling $S$ ($S/2$) and a dual gauge group $\hG$.}

Moreover, if $n$ is an integer such that $e_i+n\hm_i\in \L_W(G)$ for
any $e_i\in \L_W(G)/\L_R(G)$ and $\hm_i\in \hL_W(G)/\hL_R(G)$, then
the partition function is invariant under $\T^n$:
\beq
\T^n\,\, : \qquad \Z(G,S)\to \Z(G,S+ni)=\Z(G,S).
\eeq{TofZ}
Therefore, as it will be shown below,
the partition function is invariant
under the subgroup generated by $\T^n$ and $\S\T^{\hn}\S$. Here $\hn$ is
an integer such that $\hm_i+\hn e_i\in \hL_W(G)$ for any
$e_i\in \L_W(G)/\L_R(G)$ and $\hm_i\in \hL_W(G)/\hL_R(G)$; for such an
$\hn$ the free energy for the dual group $\hG$ is invariant under
$\T^{\hn}$:
\ber
\Z(\hG,S+\hn i)&=&\Z(\hG,S), \qquad\,\,\, {\rm if}\,\,\,\,\,
G \,\, {\rm simply-laced} \nonumber\\
\Z(\hG,(S+\hn i)/2)&=&\Z(\hG,S/2), \qquad {\rm if}\,\, G \,\, {\rm
non-simply-laced.}
\eer{ZZhG}
We thus find that
\ber
\S\T^{\hn}\S(\Z(G,S))=\S\T^{\hn}(\Z(\hG,S))=\S(\Z(\hG,S))=\Z(G,S),
& &{\rm if}\,\,\,\,\, G \,\, {\rm simply-laced} \nonumber\\
\S\T^{\hn}\S(\Z(G,S))=\S\T^{\hn}(\Z(\hG,S/2))=\S(\Z(\hG,S/2))=\Z(G,S),
& & {\rm if}\,\, G \,\, {\rm non-simply-laced.} \nonumber\\
\eer{check}
This shows that $\Z$ is invariant under the subgroup of $SL(2,Z)$ generated
by $\T^n$ and $\S\T^{\hn}\S$.
\section{Summary and Conclusions}
\setcounter{equation}{0}
In this paper we have defined some gauge invariant quantities in  $N=4$ super
Yang-Mills theories based on arbitrary compact, simple groups (the
generalization to arbitrary compact groups is straightforward).
They are the partition
functions with twisted boundary conditions and the corresponding free
energies. To define the twist, as well as the non-Abelian equivalent of the
electric and magnetic fluxes, we extended straightforwardly the definitions
given by 't Hooft for the $SU(N)$ case. The partition function is infrared
divergent, and the main result of this paper is that the leading infrared
divergence, in the expansion given in eq.~(\ref{m1}), is exactly computable.
Upon performing this computation, we derived the corresponding leading
divergence of the free energies in all flux sectors.

We defined the
transformation laws under S-duality of the free energies. These laws are
general and must hold in any theory $SL(2,Z)$-duality symmetric, and
whose transformation laws under S-duality obey the Montonen-Olive
conjecture. Finally we
verified that these laws are obeyed by the quantities we computed, thereby
providing another independent check of the S-duality conjecture in
$N=4$ supersymmetric theories.

In studying S-duality on non-simply laced groups, we noticed an interesting
phenomenon. Namely, we found in Section 6.3 that
there exist $SL(2,Z)$ transformations that are not
allowed, since they would transform physical fluxes into unphysical ones.
Obviously, one can modify the definition of S-duality so as to cure this
problem. This is done by defining the ${\cal T}$ transformation as being
always a $2\pi$ theta shift.
Thus, one defines ${\cal TS}$ as (Cfr. eqs.~(\ref{a3}) and (\ref{a2}))
\beq
{\cal TS}(F[\ve,\vhm,S])={\cal T}(F[\vhm/\sqrt{2},-\sqrt{2}\ve,S/2])\equiv
F[\vhm/\sqrt{2},-\sqrt{2}\ve,S/2 +i].
\eeq{conc1}
Even with this new definition, only a subgroup of all
fractional linear transformations $iS'= (aiS+b)/(ciS +d)$, $a,b,c,d
\in Z$, $ad-bc=1$ is realized by transformations among physical fluxes.
For instance, as we saw in Section 6.3,
$iS'=-(1/iS+1)$ does not belong to this subgroup.
This means that when
$S$ is promoted to a true dynamical field, only a subgroup of $SL(2,Z)$
becomes a true symmetry. It is interesting to notice that this reduction of
symmetry happens only in non-simply laced gauge groups,
which can never be obtained
from $N=4$ supersymmetric compactifications of the heterotic string!

Finally, we should remark that further highly non-trivial tests of
S-duality in $N=4$ Yang-Mills theories
could be done by computing the subleading terms in the infrared-divergence
expansion~(\ref{m1}).

\vskip .5in \noindent
{\bf Acknowledgements} \vskip .2in \noindent
We would like to thank S. Elitzur, C. Imbimbo, S. Mukhi, E. Rabinovici, A.
Schwimmer, C. Vafa and E. Witten for discussions.
L.G. is supported in part by the
Ministero dell' Universit\`a e della Ricerca Scientifica e Tecnologica,
by INFN and by ECC, contracts SCI*-CT92-0789 and CHRX-CT92-0035.
A.G. is supported in part by BSF - American-Israel Bi-National
Science Foundation, by the BRF - the Basic Research
Foundation and by an Alon fellowship. M.P. is supported in part by NSF under
grant no. PHY-9318171. A.Z. is supported in part by ECC, Project
ERBCHGCTT93073, SCI-CT93-0340, CHRX-CT93-0340.

\vskip .5in \noindent
{\bf Note Added} \vskip .2in \noindent
In two independent papers~\cite{new}, which appeared during the completion of
this manuscript,
a reduction of 4-$d$ super YM to supersymmetric 2-$d$ sigma models is
considered, leading to a map of S-duality to T-duality of the SCFT.
Some of the results of~\cite{new} could be related to this work,
in particular, to eqs.~(\ref{FEMsl}) and~(\ref{Zaf2'}).

\section*{Appendix A - Notations and Algebra}
\setcounter{equation}{0}
\renewcommand{\theequation}{A.\arabic{equation}}
Let $G$ denote a compact, simple Lie group, and let $\t{G}$ be the
universal covering group of $G$ (the generalization to semi-simple
groups is obvious). The center of $G$ -- the set of
elements commuting with {\em all} the group elements -- will be denoted by
$C(G)$. The center of $\t{G}$ will be denoted by $C\equiv C(\t{G})$.
The group $G$ is related to its universal covering by the quotient:
\beq
G={\t{G}\over K},\qquad K\subseteq C.
\eeq{GC}
The Lie algebra of $G$ will be denoted by $\G$.
Obviously, the Lie algebra is the same for any sub-group $K\subseteq C$,
because the center of a compact, semi-simple group is discrete.
The dual group of $G$  will be denoted by
\beq
\hG\equiv G^{dual}.
\eeq{hG}
The dual Lie algebra -- the Lie algebra of $\hG$ -- will be denoted by
\beq
\hg\equiv \G^{dual}.
\eeq{hcG}
The definition of the dual group and the dual algebra will become evident
from the discussion below. Here we shall only mention a few properties of
the duality operation. First, the dual of a dual group is the original
group:
\beq
dual^2\equiv (\wedge)^2=1.
\eeq{dd}
The dual Lie algebra equals to the original Lie algebra when the group is
simply-laced:
\beq
G\,\, {\rm simply-laced}\, \implies\, \G=\hg,
\eeq{sl}
and for non-simply-laced groups:
\beq
\G=so(2n+1)\, \Leftrightarrow\, \hg=sp(2n), \qquad
\G=\hg \,\,{\rm for}\,\, G_2,F_4.
\eeq{nsl}

Next we shall discuss several lattices: the root lattice, the weight lattice
and their duals. The root lattice of the Lie algebra $\G$ will be denoted
\beq
\L_R\equiv \L_{Root}(\G),
\eeq{LR}
and we denote a basis of the root lattice by
\beq
\{e_n\}_{n=1,...,r}\in \L_R, \qquad r={\rm rank}\, G,
\eeq{eLR}
where $e_n$ is a vector with $r$ components: $e_n\equiv (e_n^1,..,e^r_n)$.
A convenient choice for a basis is the set of simple roots,
normalized such that
\ber
\{e_n\}_{n=1,...,r}&=&\{{\rm simple\,\, roots}\}, \nonumber\\
e_n^2&=&2\qquad {\rm for\,\, long\,\, roots},\qquad  \nonumber\\
(&\implies& e_n^2=1\qquad {\rm for\,\, short\,\, roots}).
\eer{els}
In this work, we assume this choice, unless
otherwise specified. We define the $r\times r$ matrix $C$ by
\beq
C_{nm}=e_n\cdot e_m \equiv \sum_{P=1}^r e_n^P e_m^P.
\eeq{Cnm}
$C$ is related to the Cartan matrix $2(e_n\cdot e_m)/e_n^2$; the two are
equal if $\G$ is simply-laced.

The weight lattice of the Lie algebra $\G$ will be denoted
\beq
\L_W\equiv \L_{Weight}(\G),
\eeq{LW}
and we denote a basis of the weight lattice by
\beq
\{e^*_n\}_{n=1,...,r}\in \L_W .
\eeq{eLW}
The basis $e_n^*$ can be chosen such that it obeys:
\beq
{2(e_n\cdot e_m^*)\over (e_n\cdot e_n)}=\delta_{nm}.
\eeq{eeee}
For simply-laced groups, all roots have the same length ($e_n^2=2$ for all
$n$) and, therefore, (\ref{eeee}) implies that
\beq
\G\,\, {\rm simply-laced}\, \implies\, e_n^2=2\, \implies\, e_n\cdot
e_m^*=\delta_{nm}.
\eeq{eed}
However, for non-simply-laced groups there are long roots ($e_n^2=2$) as well
as short roots ($e_n^2=1$) and, therefore, (\ref{eeee}) implies that
\beq
\G\,\, {\rm non-simply-laced}\, \implies\, e_n^2=2\,\, {\rm or}\,\, 1\,
\implies\, e_n\cdot e_m^*=\delta_{nm}\,\, {\rm or}\,\, {1\over 2}\delta_{nm}.
\eeq{eedd}

The dual lattice of the weight lattice of $\G$ is equal (up to
normalization) to the root lattice of the dual algebra $\hg$ and,
therefore, we will denote it by
\beq
\hL_R \equiv (\L_W(\G))^{dual}.
\eeq{hLR}
A basis to $\hL_R$ will be denoted
\beq
\{\he_n\}_{n=1,...,r}\in \hL_R.
\eeq{he}
The basis $\he_n$ can be chosen such that it obeys:
\beq
\he_n\cdot e_m^*=\delta_{nm}.
\eeq{hee}
{}From eqs. (\ref{eeee}) and (\ref{hee}) it thus follows that
\beq
\he_n={2e_n\over (e_n\cdot e_n)},
\eeq{heee}
and from (\ref{heee}) one can show that the explicit relation between
$\hL_R$ and the root lattice of the dual algebra is:
\ber
G\,\, {\rm simply-laced}\, &\implies& \, \hL_R=\L_R , \nonumber\\
G\,\, {\rm non-simply-laced}\, &\implies& \, \hL_R=\sqrt{2}\L_R(\hg).
\eer{ehLR}

The dual lattice of the root lattice of $\G$ is equal (up to
normalization) to the weight lattice of the dual algebra $\hg$ and,
therefore, we will denote it by
\beq
\hL_W \equiv (\L_R(\G))^{dual}.
\eeq{hLW}
A basis to $\hL_W$ will be denoted
\beq
\{\he^*_n\}_{n=1,...,r}\in \hL_W.
\eeq{heW}
The basis $\he_n^*$ can be chosen such that it obeys:
\beq
\he^*_n\cdot e_m=\delta_{nm}.
\eeq{heeW}
{}From eqs. (\ref{Cnm}) and (\ref{heeW}) it follows that
\beq
\he_n^*\cdot \he_m^* = (C^{-1})_{nm}.
\eeq{Cinv}
By using the definition of $\he_n$ and $\he_m^*$ and eq. (\ref{eeee})
one can find that
\beq
{2(\he_n\cdot \he_m^*)\over (\he_n\cdot \he_n)}=\delta_{nm}
\eeq{heeeW}
{}From eqs. (\ref{heeeW}) and (\ref{ehLR}) it thus follows that
the explicit relation between
$\hL_W$ and the weight lattice of the dual algebra is:
\ber
G\,\, {\rm simply-laced}\, &\implies& \, \hL_W=\L_W , \nonumber\\
G\,\, {\rm non-simply-laced}\, &\implies& \, \hL_W=\sqrt{2}\L_W(\hg).
\eer{ehLW}

Finally, from eq. (\ref{heee}) one finds that
\ber
G\,\, {\rm simply-laced}\, &\implies& \, e_n^2=2\, \implies\,
\he_n^2=2 , \nonumber\\
G\,\, {\rm non-simply-laced}\, &\implies& \, e_n^2=2\,\, {\rm or}\,\, 1\,
\implies\, \he_n^2=2\,\, {\rm or}\,\, 4.
\eer{he24}
{}From eqs. (\ref{heeeW}) and (\ref{he24}) we thus learn that
\ber
G\,\, {\rm simply-laced}\, &\implies& \, \he_n\cdot\he_m^*=\delta_{nm} ,
\nonumber\\
G\,\, {\rm non-simply-laced}\, &\implies& \,
\he_n\cdot\he_m^*=\delta_{nm}\,\, {\rm or}\,\, 2\delta_{nm}.
\eer{GG12}

The center, $C$, of $\t{G}$ is:
\beq
C=\{e^{2\pi i \hw\cdot T} | \hw\in \hL_W/\hL_R\}.
\eeq{Center}
Here $\hw$ is a vector with components $\hw^P$, $P=1,...,r$, $r={\rm
rank}\, G$, and $\{T_P\}_{P=1,...,r}$ are the generators in the CSA.
A weight $w=(w_1,...,w_r)$ is the eigenvalue of $(T_1,...,T_r)$
corresponding to one common eigenvector in a single valued representation
of $G$:
\beq
T_PV_w=w_PV_w, \qquad w\in \L_W.
\eeq{Vw}
(We should remark that eq. (\ref{Center}) is true for any choice of
normalization, because (\ref{Vw}) implies that the normalization of
$w\in \L_W$ and $\a\in \L_R$ is compatible with the choice of generators
$T$. Thus $C$ is indeed the center, because
$w\cdot \hat{\a}\in Z$ for any $w\in \L_W$ and $\hat{\a}\in
\hL_R=(\L_W)^{dual}$, and this can be used to prove that the elements in
(\ref{Center}) are the full set of group elements which
commute with all the generators.)

For convenience let us summarize our notations and normalizations:
\begin{itemize}
\item
$G$ is a compact, simple group.
\item
$\t{G}$ is the universal covering group of $G$.
\item
$G=\t{G}/K$, $K\subseteq C$, $C\equiv Center(\t{G})$.
\item
$\G$ is the Lie algebra of $G$.
\item
$\hG$ is the dual group of $G$.
\item
$\hg$ is the dual Lie algebra, {\em i.e.}, the Lie algebra of $\hG$.
\item
$\L_R$ is the root lattice of $\G$, with normalization $({\rm long\,\,
root})^2=2$.
\item
$\L_W$ is the weight lattice of $\G$.
\item
$\hL_R$ is the dual lattice of $\L_W$.
\item
$\hL_W$ is the dual lattice of $\L_R$.
\item
$\hL(\G)=N(\G)\L(\hg)$, where $N(\G)=1$ if $G$ is simply-laced, and
$N(\G)=\sqrt{2}$ if $G$ is non-simply-laced; $\L$ is either $\L_R$ or
$\L_W$.
\item
The group $G$ has a weight lattice of representations which is a
sub-lattice of $\L_W$: $G=\t{G}/K$ $\implies$ $\L_W(G)=\L_W/K$.
\item
The dual group $\hG$ has a weight lattice dual to the weight lattice of
$G$: $\L_W(\hG)=\L_W(G)^{dual}$.
\item
$\hL(G)_{R,W}=N(\G)\L(\hG)_{R,W}$, where $N(\G)=1$ if $G$ is simply-laced, and
$N(\G)=\sqrt{2}$ if $G$ is non-simply-laced.
\item
For $G$ simply-laced: $\hg=\G$.
\item
For $G$ non-simply-laced: $\G=so(2n+1)\, \Leftrightarrow\, \hg=sp(2n)$.
(The Lie algebrae of $G_2$ and $F_4$ are self-dual.)
\end{itemize}

\section*{Appendix B - the Hamiltonian Formalism}
\setcounter{equation}{0}
\renewcommand{\theequation}{B.\arabic{equation}}
The free energy $F[\ve,\vhm,S]$ can be calculated, in principle,
also in the Hamiltonian
formalism~\cite{GGPZ}, and one can check that our result is consistent with the
weak coupling Hamiltonian evaluation. The free
energy at the different flux sectors reads~\cite{tH}:
\beq
e^{-\b F[\ve,\vhm,S]}={\rm Tr}_{\vhm} P[\ve]e^{-\b H} ,
\eeq{FH}
\beq
P[\ve]=\prod_{i=1}^3 P[e_i], \qquad P[e_i]={1\over N} \sum_{\hk_i\in
\hL_W/\hL_R} \O_{\hk_i} e^{2\pi i \hk_i\cdot e_i}
\eeq{Pe}
($N=Order(C)$),
\beq
\O_{\hk_i}=e^{2\pi i \hk_i\cdot T {x_i\over a_i}}
\eeq{Om}
(the generators $T$ are in the CSA, see Appendix A).
The symbol ${\rm Tr}_{\vhm}$ in (\ref{FH}) denotes the trace over gauge
fields obeying boundary conditions twisted by $\vhm$. $P[\ve]$ is the
projector onto states with a definite value of the electric fluxes. The
gauge transformations $\O_{\hk_i}$ in eq. (\ref{Om}) are periodic up to a
non-trivial element of the center of $\t{G}$, $\exp{(2\pi i \hk_i\cdot T)}\in
C$ :
\ber
\O_{\hk_i}(x_j+a_j)&=&e^{2\pi i \hk_i\cdot T} \O_{\hk_i}(x_j) \qquad {\rm if}
\,\,\, j=i, \nonumber\\
\O_{\hk_i}(x_j+a_j)&=& \O_{\hk_i}(x_j) \qquad\qquad\,\,\,\,\,
{\rm if} \,\,\, j\neq i .
\eer{Oxa}

It must be stressed that the free energy in eq. (\ref{FH})
is the finite temperature
free energy which assumes antiperiodicity in time for the fermions.
Unfortunately, it is complicated to evaluate this trace for every regime.
But, as in \cite{GGPZ},
one can compute its $\theta=\vhm=0$, $g\ll 1$ limit, finding complete
agreement with the leading infrared-divergent part of the free energy,
evaluated
with the functional integral approach in Section 4. The role of the infrared
divergence is crucial to
get the correct result for a general compact group
\cite{imbimbo}.\footnote{
We thank C. Imbimbo and S. Mukhi for informing us about their results prior to
publication.}

Since the coupling constant is small, we
can compute $F[\ve,0,g]$ by using perturbation theory. This possibility
exists since $N=4$ super Yang-Mills has a vanishing beta function;
the theory with $g\ll 1$ is weakly interacting at {\em any}
length scale. We set $\vhm$ to zero since in this regime the magnetic flux
is expected to interact strongly, with coupling constant $O(4\pi/g)\gg 1$.
At $\vhm=0$ all fields in the box obey periodic boundary conditions, up
to a periodic, globally defined gauge transformation. In the box,
all their non-zero Fourier modes have energies $O(V^{-1/3})$. Moreover,
the potential energy scales as $1/g^2$. For instance, the gauge fields
possess a ``magnetic'' energy $\int d^3 x g^{-2}\bf{B}^2$.
The zero modes belonging to the
Cartan subalgebra, instead, have energies $O(g^2V^{-1/3})$.
At $g\ll 1$ only this latter set gives a significant contribution to the
free energy~\cite{WI,l}. If the spatial supersymmetry is enforced by choosing
periodic spatial boundary conditions for fermions, the quantum fluctuations
do not destroy the classical degeneracy of the manifold of the gauge
zero modes, unlike in the non-supersymmetric case~\cite{l}. The low
energy spectrum is therefore obtained by quantizing such manifold. The
Hamiltonian
evaluation with periodic fermions (in space) continues to be a good check at
weak coupling of our result since
this different choice of boundary conditions does not
affect the functional integral evaluation of Section 4.

The configurations contributing to the partition function at $g\ll 1$ are,
therefore, the general gauge transformations
of the zero modes belonging to the CSA.
Notice that the constant gauge configurations, $\bf{c}$, are
defined only up to periodic gauge transformations. The element
\beq
\O_{\hat r_i}=e^{2\pi i \hat r_i\cdot T {x_i\over a_i}}, \qquad
\hat r_i\in\hat\Lambda_R , \qquad i=1,2,3
\eeq{Zaf1}
is well-defined and periodic
in the box in which we have defined the system; thus it generates a gauge
transformation, which acts on the zero modes as
\beq
c_i \rightarrow c_i + {2\pi\over a_i}\hat r_i .
\eeq{Zaf2}
Therefore, the zero modes $c_i$ are periodic; moreover, we have to mod
out by the
Weyl group which is realized by other constant gauge transformations. The
only periodic gauge transformations that map the Cartan torus to itself are
$\O_{\hat r_i}$ (\ref{Zaf1}) and the constant Weyl transformations.
Thus the zero modes of the gauge fields parametrize a
compact space with boundary, called the toron manifold.
Moreover, the Weyl group acts also on constant scalar fields modes in the
CSA, $\phi_{IJ}$, thus $(\phi_{IJ},c_i)$ live on the orbifold
\beq
(\phi_{IJ},c_i)\in {R^{6r}\times \prod_{i=1}^3 T_i^r \over {\rm
Weyl-Group}}, \qquad T_i^r\equiv {R^r\over 2\pi\hL_R/a_i}, \qquad r={\rm
rank}\, G .
\eeq{Zaf2'}

The Lagrangian restricted to the modes constant in space reads
\beq
L = {V\over g^2}\int dt (\sum_i\dot c_i\cdot\dot c_i +
\dot a_I^A\cdot a_I^{A\,\dagger} + \dot\phi_{IJ}\cdot\dot\phi_{IJ}),
\eeq{Zaf3}
where $c_i,a_I^A,\phi_{IJ}$ are the zero modes in the CSA
of the gauge fields, fermions and scalars, respectively, and $A=1,2$ is the
spinor index.
In terms of the conjugate momenta, $\pi_i$, of the gauge zero modes $c_i$,
the Hamiltonian is
\beq
H = {g^2\over 4V}\sum_i\pi_i\pi_i + scalars .
\eeq{Zaf5}
Clearly, the eigenfunctions are plane waves,
\beq
e^{ik_i\cdot c_i}
\eeq{Zaf6}
(where $k_i\in a_i\Lambda_W$ for ensuring single-valuedness on the compact
space of the $c_i$), with energy,
\beq
E =
{g^2\over 4V}\sum_i k_i^2 .
\eeq{Zaf7}
The fermion zero modes do not contribute to the energy, while the scalar
zero modes contribute with the energy of a plane wave of continuous momentum.

Due to eq. (\ref{Zaf2'}),
the partition function is not simply the sum of the statistical weights
associated with these energies over all the weight lattice (i.e. it is not
a theta function),
since we have to mod out by the Weyl group. The energy term
$(g^2/4V){\bf{k}}^2$,
being the natural Cartan scalar product on the weight lattice, is invariant,
but the multiplicity with which we must count the single vector $\bf{k}$
depends on the number of invariants under the Weyl group we can construct
with the product of the
gauge, fermionic and scalar wavefunctions. Since the Weyl group does not act
freely on the weight lattice, the correct partition function,
in general, is not a theta function (in particular in the purely bosonic case).
The exact computation of the multiplicities is
a complicated group theory problem; fortunately, the existence of the
divergence due to the scalars simplifies considerably
the problem~\cite{imbimbo}.

We are to compute the partition function,
\beq
{\rm Tr}\, Pe^{-\beta H},
\eeq{Zaf8}
on the Hilbert space constructed out of the vacuum with the 8
fermionic oscillators $a_I^{A\,\dagger}$,
the gauge exponential wavefunctions
$e^{ik_i\cdot c_i}$ and the 6 scalar zero modes $e^{ip_{IJ}\cdot\phi_{IJ}}$.
The states $|k_i,a_I^A,p_{IJ}\rangle,\null k_i\in a_i\Lambda_W,\null p_{IJ}\in
R^r$ provide a suitable Fock
basis of eigenvalues of the Hamiltonian with energy
$(g^2/4V)({\bf{k}}^2+\sum_{IJ}p_{IJ}^2)$.
$P=(1/|W|)\sum_{g\epsilon W}g$ is the projector over Weyl invariant
states. The Weyl group
is generated by the reflections about simple roots
\beq
k\rightarrow k_i - (e_n\cdot k_i)e_n .
\eeq{Zaf9}
It takes the state  $|k_i\rangle =e^{ik_i\cdot c_i}$ into $|k^g_i\rangle$,
another state in the Fock
space. It leaves invariant the states with $e_n\cdot k_i=0$, i.e. the walls
of the Weyl chambers. The partition function reads
\beq
{\rm Tr}\, Pe^{-\beta H} =
{1\over |W|}\sum_{g\in W}\sum_{{\bf{k}},a_I^A}e^{-(\beta g^2/4V){\bf{k}}^2}
\prod_{IJ}\int dp_{IJ}
\langle k_i,a_I^A,p_{IJ}|g|k_i,a_I^A,p_{IJ}\rangle
e^{-(\beta g^2/4V)p_{IJ}^2} .
\eeq{Zaf9'}
Now, the matrix element $\langle k_i|g|k_i\rangle$ is different from zero only
if $k_i$ is left invariant by $g$;
so we can restrict the sum over $g$ to the little group $W_{\bf{k}}$
of $\bf{k}$ (the set of elements of $W$ which leaves $\bf{k}$ fixed):
\beq
{1\over |W|}\sum_{k_i\in a_i\Lambda_W}e^{-(\beta g^2/4V) {\bf{k}}^2}
\sum_{g\in W_{\bf{k}}}
\sum_{a_I^A}\prod_{IJ}\int dp_{IJ}
\langle a_I^A,p_{IJ}|g|a_I^A,p_{IJ}\rangle e^{-(\beta g^2/4V)
p_{IJ}^2} .
\eeq{Zaf10}
Since $W$ acts freely in the interior of the Weyl chamber, $W_{\bf{k}}$
is trivial
for such elements, becoming larger and larger on the various walls, walls
of walls etc., and becomes the complete Weyl group only at $\bf{k}=0$.

Let us compute the matrix element $\langle a_I^A,p_{IJ}|g|a_I^A,p_{IJ}\rangle$.
On the fermionic Fock space, $g$ acts as a linear transformation, thus giving
a factor
$\det (1+g)$ for each spinorial degree of freedom.
The scalars contribution is
\beq
\langle p_{IJ}|g|p_{IJ}\rangle =
\prod_{IJ}\int dp_{IJ}\int_0^R dx_{IJ} e^{-(\beta g^2/4V)[ p_{IJ}^2
+ip_{IJ}(x_{IJ}-x^g_{IJ})]} ,
\eeq{Zaf11}
where we have put a cutoff $R$ to regularize the configuration space
divergence of the scalars; whenever possible we will send $R\rightarrow\infty$.
For each scalar degree of freedom, we find
\beq
\int_0^R dx e^{-{1\over 4\beta}x(1-g)^2x}
= {R^{n_g}\over \det'(1-g)};
\eeq{Zaf12}
$n_g$ is the number of eigenvalues of $g$ equal to 1, and the prime means the
exclusion of the zero eigenvalues.

To sum up, the total partition function reads
\beq
\sum_{k\in a_i\Lambda_W}e^{-(\beta g^2/4V){\bf{k}}^2}{1\over |W|}
\sum_{g\epsilon W_{\bf{k}}}R^{n_g}{\det^8(1+g)\over \det'^6(1-g)}.
\eeq{Zaf13}
Now, it is clearly difficult to evaluate exactly the coefficients. However,
we need to take in this sum only the most divergent contribution in $R$,
since the subleading terms disappear when we normalize to get physical
quantities. For each $W_{\bf{k}}$ there is only one element with the
maximum power $n_g$ (and so with the maximum divergence): the identity. All
the other elements contribute subleading terms. We see that the most
divergent coefficient is always the same and
henceforth the exact theta function,
\beq
\sum_{k_i\in a_i\Lambda_W}e^{-\sum_i(\beta g^2/4V)k_i^2} ,
\eeq{theta}
gets reconstructed, up to a multiplicative factor.

The full partition function splits into flux sector according to the
projector (\ref{Pe}); one can check that this restricts the sum over
weight vectors exactly according to the definition given in Section 3.
This argument shows that the $\theta=\vhm=0$, $g\ll 1$ approximation of
the free energy completely agrees with the results of this paper.

Finally, let us remark that by using factorization, 't Hooft's duality,
and the Witten phenomenon (see Section 5), one can derive the
result (\ref{FS}) from the $\theta=\vhm=0$ free energy.

\end{document}